\documentclass[12pt]{article}
\usepackage{amsmath, amssymb, amsthm}
\usepackage{geometry}
\usepackage{natbib}
\usepackage{graphicx}
\usepackage{setspace}
\usepackage{tikz} 						
\usetikzlibrary{patterns,arrows.meta}
\usepackage[colorlinks, urlcolor=blue, linkcolor=blue,citecolor=blue,backref=page]{hyperref}
\usepackage{comment}
\usepackage{threeparttable}
\usepackage{booktabs}
\theoremstyle{plain}
\newtheorem{theorem}{Theorem}

\newtheorem{assumption}{Assumption}
\newtheorem{definition}{Definition}

\usetikzlibrary{patterns}

\usepackage{natbib}
\setcitestyle{sort&compress}

\AtBeginDocument{}

\hypersetup{
    colorlinks,
    linkcolor={blue!75!black},
    urlcolor={blue!75!black},
    citecolor={blue!75!black},
}

\title{Multivariate Ordered Discrete Response Models with Lattice Structures}
\author{Tatiana Komarova\thanks{Faculty of Economics, University of Cambridge. Email: tk670@cam.ac.uk} \and William Matcham\thanks{Department of Economics, Royal Holloway University of London. Email: william.matcham@rhul.ac.uk}}
\date{November 4, 2025}
\begin{document}

\maketitle

\begin{abstract}
We analyze multivariate ordered discrete response models with a lattice structure, modeling decision makers who narrowly bracket choices across multiple dimensions. These models map latent continuous processes into discrete responses using functionally independent decision thresholds. In a semiparametric framework, we model latent processes as sums of covariate indices and unobserved errors, deriving conditions for identifying parameters, thresholds, and the joint cumulative distribution function of errors. For the parametric bivariate probit case, we separately derive identification of regression parameters and thresholds, and the correlation parameter, with the latter requiring additional covariate conditions. We outline estimation approaches for semiparametric and parametric models and present simulations illustrating the performance of estimators for lattice models.

\textbf{Keywords}: Ordered response, lattice structure, semiparametric models, parametric identification, narrow bracketing

\textbf{JEL Classification}: C14, C31, C35
\end{abstract}

\section{Introduction}
Ordered response models are fundamental in empirical economics, used to analyze discrete choices with inherent ordering, such as risk aversion \citep{MalmendierNagel2011}, political violence \citep{BesleyPersson2011}, or educational attainment \citep{CameronHeckman1998}. These models map latent continuous variables to discrete outcomes via thresholds. While univariate models are well-established (\cite{CunhaHeckmanNavarro2007}, among many others), multivariate extensions can allow researchers to capture joint decisions across multiple dimensions.

Univariate ordered response models, like ordered probit and logit, were formalized by \citet{McKelveyZavoina1975} and \citet{AndresonPhilips1981}. Multivariate extensions, introduced by \citet{AshfordSowden1970} for bivariate probit models, account for correlated decisions. Psychometrics and structural-equation literature adopted and extended the latent-variable viewpoint to multiple categorical indicators. In particular, \cite{Muthen1984} formalized a structural equation framework that allowed dichotomous and ordered categorical indicators to be treated as manifestations of underlying (multivariate normal) latent variables -- effectively a multivariate ordered model within the SEM tradition. SEM/psychometrics work (e.g., \cite{Olsson1979}, among others) set out approaches for polychoric/probit models for multiple ordinal indicators. \cite{Kim1995} explicitly proposed and implemented a bivariate cumulative probit regression model for ordered categorical margins, with application and numerical estimation details. For detailed coverage of various types of univariate ordered response model we refer the reader to \cite{Agresti1990}, \cite{Boes2006}, \cite{STEWART2005}, \cite{greene2010}. \cite{greene2010} includes a review of recent applications of the bivariate ordered probit model. Applications of trivariate ordered probit models include \cite*{buliung2007,genius2005,scott2001}.

We focus on a particular class of multivariate ordered response models with a lattice structure, where decision makers narrowly bracket their choices, treating dimensions in isolation, in line with the behavioral economics framework of narrow bracketing \citep{ReadLoewensteinRabin1999}. The lattice structure is characterized by functionally independent decision thresholds across dimensions, producing a grid-like latent space -- hence our terminology `lattice models.''\footnote{This terminology is our own and is not standard in the literature.} In practice, lattice models (often coupled with some parametric assumptions on the distribution of unobservables) have often served as the default and most straightforward extension of univariate ordered response models in applied work. \cite{kmnonlattice} explicitly adopt the term ``lattice models'' to distinguish these restricted structures from more general multivariate formulations.

In this paper, we develop a formal and rigorous \emph{semiparametric framework} for lattice ordered response models, where latent processes are specified as linear combinations of covariates and unobserved errors. We derive identification conditions for regression parameters, thresholds, and the joint distribution of the unobservables. The literature on univariate ordered models models provides several foundational insights that aid some identification results in multivariate settings  as narrow bracketing allows us to isolate decision making across different dimensions. Namley, under full independence between unobservables and covariates, identification of index parameters and thresholds can rely on single-index methodologies just like in the univariate case. More general semiparametric approaches have allowed weaker conditions: \cite{lee1992} studied median independence following \cite{manski1975,manski1985,manski1988} work on maximum score, while \cite{lewbel2000} and \cite{chenkhan2003} allowed heteroskedastic unobservables with the latter focusing on multiplicative heteroskedasticity. In our analysis, we maintain full stochastic independence between unobservables and covariates to primarily focus on the identification and estimation of the joint cumulative distribution function (c.d.f.) which is a topic largely unexplored in the literature, even for lattice models.

Identification of the joint c.d.f. of unobservables in semiparametric models is a core theoretical contribution of this paper. Understanding this joint distribution is crucial for policy analysis. In lattice models, complementarity and substitutability in decision structures are not directly modeled. Thus, all dependence in observed decisions (conditional on covariates) is captured by the dependence among unobservables. This dependence structure is central to policy design involving joint outcomes such as household decisions on healthcare and education investments where the correlation between latent factors determines whether bundled interventions reinforce or crowd out each other. Semiparametric identification avoids restrictive parametric assumptions (e.g., joint normality) that can distort estimated policy effects if misspecified \citep{MalmendierNagel2011}.

From an estimation perspective, we outline how existing semiparametric estimation methods can recover index parameters and thresholds in a  sample, and we discuss how one could approach the estimation of the joint c.d.f. after those parameters are estimated at the $\sqrt{n}$-rate. We also describe how the approach of \cite{Coppejans2007} can be extended to jointly estimate all unknown components in one step.

For the parametric case, we focus on the multivariate normal specification, which conveniently captures varying degrees of dependence.\footnote{Alternative parametric specifications for the joint c.d.f. in bivariate ordered response models include \cite{Forcina2008} and \cite{Ferdous2010}.} Because the lattice structure allows identification results for thresholds and indices to extend from univariate models, our attention centers on identifying the correlation parameters. We provide several sufficient conditions for identification in the bivariate case, including (i) configurations where one latent index is pinned at zero, (ii) variation in sign of index–threshold differences across subgroups, and (iii) the presence of exclusive covariates that shift one margin but not the other.

In short, this paper provides a rigorous foundation for lattice ordered response models, establishing semiparametric identification and outlining estimation strategies that make these models suitable for empirical applications where narrow bracketing is plausible such as consumer preference formation \citep{Train2009} and policy evaluation \citep{HeckmanVytlacil2007}.

The remainder of the paper is structured as follows. Section \ref{sec:model} introduces the general multivariate lattice model. Section \ref{sec:semiparam} develops the semiparametric specification, identification results and also discusses various approaches to estimation includiing those that utilize existing estimation techniques for univariate models, Section \ref{sec:param} details the parametric model focusing on multivariate normal errors and identification of correlation coefficients. Section \ref{sec:montecarlo} presents simulation evidence, and Section \ref{sec:appl} provides an empirical application estimating a joint ordered response model for health and happiness rankings. Section \ref{sec:concl} concludes. The Appendix collects proofs of the main theoretical results.

\section{Model formulation}
\label{sec:model}

We model a single agent’s decisions across $D \geq 2$ dimensions, mapping a $D$-variate latent continuous metric $\left(Y^{* c_1}, \ldots, Y^{* c_D}\right)$ to a discrete metric $\left(Y^{c_1}, \ldots, Y^{c_D}\right)$. Discrete responses in dimension $d$ are $y_j^{(d)}, j=1, \ldots, M_d$, with ordering $y_1^{(d)} < \ldots < y_{M_d}^{(d)}$.

\begin{definition}[Lattice Model]
A multivariate ordered discrete response model is a lattice model if
\[
\left(Y^{c_1}, \ldots, Y^{c_D}\right) = \left(y_{j_1}^{(1)}, \ldots, y_{j_D}^{(D)}\right) \Longleftrightarrow Y^{* c_d} \in \mathcal{I}_{j_d}^{(d)} \equiv \left(\alpha_{j_d-1}^{(d)}, \alpha_{j_d}^{(d)}\right] \quad \forall d=1, \ldots, D,
\]
with threshold normalizations
\[
\forall d=1, \ldots, D, \quad \alpha_{j_d}^{(d)} = +\infty \text{ when } j_d = M_d, \quad \alpha_{j_d}^{(d)} = -\infty \text{ when } j_d = 0.
\]
\end{definition}

Thresholds $\alpha_{j_d}^{(d)}$ depend only on $j_d$, ensuring functionally independent decision rules across dimensions. The intersections of these threshodls across different dimensions form a lattice in $\mathbb{R}^D$. This reflects narrow bracketing \citep{ReadLoewensteinRabin1999}, with intervals $\mathcal{I}_{j_d}^{(d)}$ partitioning $\mathbb{R}$ and rectangles $\times_{d=1}^D \mathcal{I}_{j_d}^{(d)}$ partitioning the latent space.

\section{Semiparametric specification}
\label{sec:semiparam}

The $d^{\text{th}}$ latent process is
\[
Y^{* c_d} = x_d \beta_d + \varepsilon_d, \quad d=1, \ldots, D,
\]
where $x_d$ is a row vector of covariates, $\beta_d$ a column vector of parameters, and $\varepsilon_d$ an error term. Errors in $\left(\varepsilon_1, \ldots, \varepsilon_D\right)$ may be correlated, allowing latent processes $Y^{*cd}$ to be correlated conditional on observables.

Let $x = (x_1, \dots, x_D)$ and $\varepsilon = (\varepsilon_1, \dots, \varepsilon_D)'$ combine full vectors of covariates and unobservables, respectively. Denote the joint c.d.f. of $\varepsilon$ as $F$ and the marginal c.d.f. of $\varepsilon_d$ as $F_d$, $d=1,\dots,D$. The length of vector $x_d$ is $k_d$, $d=1,\dots,D$. Let $\mathcal{X}_d$ denote the support of $x_d$ and for each $d$, define
\[
S^{(d;j)} = \{x_d \in \mathcal{X}_d \mid P(Y^{(d)} \leq y^{(d)}_j | x_d) \in (0,1)\}, \quad j=1, \ldots, M_d,
\]
and 
$S^{(d)} = \cup_{j=1}^{M_d} S^{(d;j)}$. Let $x_{d,m}$ denote the $m$th component of $x_d$ and $x_{d,-m}$ denote the subvector of $x_d$ excluding the $m$th component, with similar notations for $\beta$. $S^{(d)}_m$ denotes the projection of $S^{(d)}$ on $x_{d,m}$ with $S^{(d)}_{-m}$ being the projection on $x_{d,-m}$.

\subsection{Identification}
We derive identification conditions for $\beta_d$, thresholds $\alpha_{j_d}^{(d)}$ and the joint c.d.f. of unobservables under certainm assumptions. We start with Assumption  \ref{assn:ind}.

\begin{assumption}
\label{assn:ind}
For all $d=1, \ldots, D$, $\varepsilon_d$ is independent of $x_d$ and has a convex support.
\end{assumption}

 In univariate ordered response models, the assumption of independence between the unobservable and covariates is common, being used in \cite*{klein2002}, \cite*{Coppejans2007}, among many others.\footnote{Some papers (see e.g. \cite{chenkhan2003}) on univariate ordered response allow for heteroskedasticity. In our framework, this would correspond to $\sigma_d(x_d,\theta_0) \varepsilon_d$ with independent $\varepsilon_d$. Some other papers further deviate from the setting of independence. \cite*{lee1992} considers ordered response under the median independence assumption from \cite{manski1975, manski1985}. In a recent paper, \cite*{wangchen} take a partial identification approach and consider a generalized maximum score estimator when regressors are interval measured. All of these settings are beyond the scope of this paper and provide interesting avenues for extensions of our work.} We formulate an analogue of a rank condition in the form of Assumption \ref{assn:rank_ind_semiparametric}. 
\begin{assumption} 
\label{assn:rank_ind_semiparametric} 
$S^{(d)}$ is not contained in any proper linear subspace of $\mathbb{R}^{k_d}$ and $P\left(S^{(d)}\right)>0$, for any $d=1,\ldots,D$. 
\end{assumption}

\begin{theorem}
\label{th:semiparamident1}
Suppose Assumptions \ref{assn:ind},  \ref{assn:rank_ind_semiparametric} hold and for each $d=1, \ldots, D$, for some $j=1,\ldots, M_d-1$ the set $ S^{(d;j)}$ contains $\widetilde{S}^{(d;j)} = (\underline{x}_{d,1}, \overline{x}_{d,1}) \times \widetilde{S}^{(d;j)}_{-1}$ where $\underline{x}_{d,1}< \overline{x}_{d,1}$ and $\widetilde{S}^{(d;j)}_{-1}$ is not contained in any proper linear subspace of $\mathbf{R}_{K_d-1}$ and $P(\widetilde{S}^{(d;j)}_{-1})>0$. In addition, suppose $\beta_{d,1}\neq 0$.  Then, $\beta_d$ are identified up to scale.\footnote{For notational simplicity we supposed that it is the first covariate that varies within an interval and has a non-trivial impact within dimension $d$. This is without a loss of generality and generally it  can be  some other covariate $x_{d,m(d)}$ with such properties.}
\end{theorem}

Identification of threshold differences or gaps requires additional conditions to those assumed in Theorem \ref{th:semiparamident1}. This is given in Theorem \ref{th:semiparamident2}.

\begin{theorem}
\label{th:semiparamident2}
Suppose for a given $d$ conditions of Theorem \ref{th:semiparamident1} hold for any $j=1,\ldots,M_d-1$. Also, for any $j=1,\ldots,M_d-2$, there is a positive measure of $x_d \in S^{(d;j)}$ such that 
$$P\left(Y^{c_d} \leq y^{(d)}_{j} \, | \, x_d \right) = P\left( Y^{c_d} \leq y^{(d)}_{j+1} \, | \, \tilde{x}_{d}\right)$$
for some $\tilde{x}_d \in S^{(d;j+1)}$. Then $\alpha^{(d)}_{j+1}-\alpha^{(d)}_j$ is identified, $j=1,\ldots,M_d-2$. 
\end{theorem}

The new condition of Theorem \ref{th:semiparamident2} would be guaranteed if for sets $S^{(d;j)}$ and $S^{(d;j+1)}$ the intersection of the sets of probabilities $\left\{P\left(Y^{c_d} \leq y^{(d)}_{j} \, | \, x_d \right): x_d \in  S^{(d;j)} \right\}$ and $\left\{P\left(Y^{c_d} \leq y^{(d)}_{j+1} \, | \, x_d \right): x_d \in  S^{(d;j+1)} \right\}$ contains an interval $(\underline{p}_j,\overline{p}_{j})$.  Large support conditions would, e.g, ensure that this interval is $(0,1)$.

Figure \ref{fig:latticeid_ex}, which shows a bivariate lattice model, presents an intuitive summary of the identification strategy in the models with lattice structures. We consider each dimension individually and, within that dimension, express probabilities of discrete values up to certain points in terms of the marginal c.d.f. of the unobservable in that dimension and the index in that dimension. Theorem \ref{th:semiparamident1} is based on consideting just one shaded area for many different $x_d$ -- either the one the left panel or the one on the right panel in Figure \ref{fig:latticeid_ex}. Theorem \ref{th:semiparamident2} requires the computation of both shaded regions for many   different $x_d$. 

\begin{figure}[!t]
\centering
\caption{Intuition for lattice model identification}
\begin{tikzpicture}[scale=0.25]

\fill [opacity = 0.3, pattern=north west lines, pattern color=blue] (-3,9.5) rectangle (-9.5,-9.5);
\draw [very thick] (-3,-9) -- (-3,9);
\draw [very thick] (3,-9) -- (3,9);

\draw [very thick] (-9,-3) -- (9,-3);
\draw [very thick] (-9,3) -- (9,3);

\node at (-3,-3) [circle,fill,inner sep=1.5pt]{};
\node at (3,3) [circle,fill,inner sep=1.5pt]{};

\node at (-3,3) [circle,fill,inner sep=1.5pt]{};
\node at (3,-3) [circle,fill,inner sep=1.5pt]{};


\node [right, thick ] at (-3,-9) {\large $\alpha^{(1)}_1$};
\node [right, thick ] at (3,-9) {\large $\alpha^{(1)}_2$};
\node [above, thick ] at (9,-3) {\large $\alpha^{(2)}_1$};
\node [above, thick ] at (9, 3) {\large $\alpha^{(2)}_2$};
\end{tikzpicture}
\hskip 0.3in 
\begin{tikzpicture}[scale=0.25]
\fill [opacity = 0.3, pattern=crosshatch, pattern color=orange] (3,9) rectangle (-9,-9);
\draw [very thick] (-3,-9) -- (-3,9);
\draw [very thick] (3,-9) -- (3,9);

\draw [very thick] (-9,-3) -- (9,-3);
\draw [very thick] (-9,3) -- (9,3);

\node at (-3,-3) [circle,fill,inner sep=1.5pt]{};
\node at (3,3) [circle,fill,inner sep=1.5pt]{};

\node at (-3,3) [circle,fill,inner sep=1.5pt]{};
\node at (3,-3) [circle,fill,inner sep=1.5pt]{};


\node [right, thick ] at (-3,-9) {\large $\alpha^{(1)}_1$};
\node [right, thick ] at (3,-9) {\large $\alpha^{(1)}_2$};
\node [above, thick ] at (9,-3) {\large $\alpha^{(2)}_1$};
\node [above, thick ] at (9, 3) {\large $\alpha^{(2)}_2$};
\end{tikzpicture}
\caption*{Notes: Left region in the latent space corresponds to $P\left(Y^{(1)}\leq y_1^{(1)}|x_1\right)$. Right region corresponds to $P\left(Y^{(1)}\leq y_2^{(1)} | x_1\right)$.}
\label{fig:latticeid_ex}
\end{figure}

The result of Theorem \ref{th:semiparamident2} immediately implies conditions for identification of marginal distributions of $\varepsilon_d$, $d=1, \ldots, D$. 

\begin{theorem}\label{corollary:Fd}
	Suppose conditions of Theorem \ref{th:semiparamident2} hold for some $d$. Suppose that 
    \begin{equation}
    \label{extracond}  \bigcup\limits_{j=1,\ldots, M_d-1} \bigcup\limits_{x_d \in S^{(d;j)}} P\left(Y^{(d)}\leq y_j^{(d)}|x_d\right)=(0,1).
    \end{equation}

    Then $F_d(\cdot)$ is identified  if (i) either one of the thresholds  among  $\alpha^{(d)}_{j}$, $j=1,\ldots, M_d-1$, is normalized to a known value, or (ii) if there is a normalization of one of the values of c.d.f. $F_d$, say
    $F_d(e_{0d})=c_{0d}$,  
    for some known $e_{0d}$ in the support of $\varepsilon_d$ and some known $c_{0d} \in (0,1)$.

\end{theorem}	
Condition (\ref{extracond}) ensures that any point in the support of $\varepsilon_d$ corresponds to the underlying $\alpha^{(d)}_j-x_d\beta_d$  for some $j$ and $x_d$. Condition (i) explicitly normalizes one threshold (the identification of values of the other thresholds then immediately follows from Theorem \ref{th:semiparamident2}), whereas condition (ii) enforces a normalization of one threshold in an indirect way.

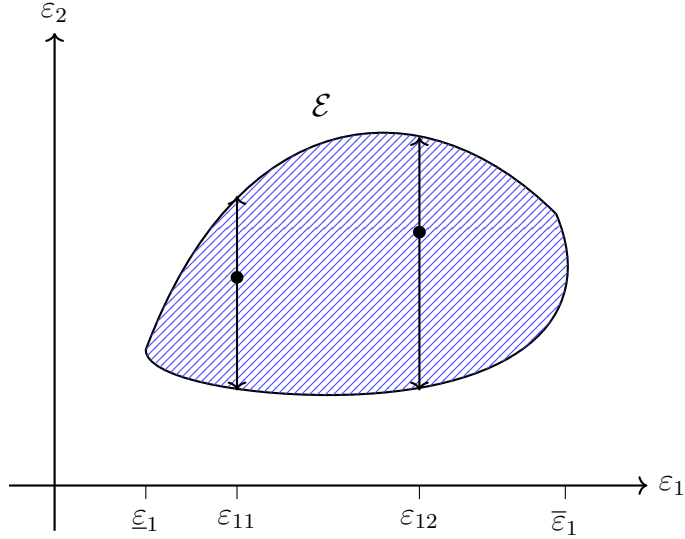
\begin{figure}
\centering
\begin{tikzpicture}[scale=1.2]

\draw[->,thick] (-0.5,0) -- (6.5,0) node[right] {$\varepsilon_1$};
\draw[->,thick] (0,-0.5) -- (0,5) node[above] {$\varepsilon_2$};

\draw[thick] (1,1.5) 
  .. controls (2,4.2) and (4,4.5) .. (5.5,3)   
  .. controls (6,1.8) and (5,1) .. (3,1)       
  .. controls (2,1) and (1,1.2) .. (1,1.5)
  -- cycle;

\path[pattern=north east lines,pattern color=blue!70] 
     (1,1.5) 
  .. controls (2,4.2) and (4,4.5) .. (5.5,3)
  .. controls (6,1.8) and (5,1) .. (3,1)
  .. controls (2,1) and (1,1.2) .. (1,1.5)
  -- cycle;

\draw[<->,thick] (2,1.05) -- (2,3.2);
\draw[<->,thick] (4,1.05) -- (4,3.85);

\fill (2,2.3) circle (2pt);
\fill (4,2.8) circle (2pt);

\foreach \x/\lbl in {1/{$   \underline{\varepsilon}_{1}$}, 2/{$\varepsilon_{11}$}, 4/{$\varepsilon_{12}$}, 5.6/{$\overline{\varepsilon}_{1}$}}{
    \draw (\x,0) -- (\x,-0.15) node[below] {\lbl};
}

\draw (2.7,4.2) node[right] {$\mathcal{E}$};

\end{tikzpicture}
\caption{ Illustration of the identification of joint c.d.f. of $\varepsilon$}
\label{idCDF}
\end{figure}

The result of Theorem \ref{corollary:Fd} does not guarantee identification of the joint distribution of unobservables, even if the conditions of this corollary hold for every $d=1,\ldots,D$. The reason is two-fold. First, Assumption \ref{assn:ind}  does not give any information about how the vector $\varepsilon$ relates to  $x_h$, $h \neq d$. Under a full stochastic independence of the vector $\varepsilon$ from the whole vector $x$, the identification process easier as $P(\varepsilon_1 \leq e_1, \ldots, \varepsilon_D \leq e_D|x)$ does not depend on $x$ and we only need to identify one $D$-variate c.d.f.   $F(e_1,\ldots, e_D)=P(\varepsilon_1 \leq e_1, \ldots, \varepsilon_D \leq e_D)$. The main channel in which we can proceed with identification of $F$ is considering observed probabilities 
$$P\left(Y^{(1)} \leq y^{(1)}_{j_1}, \ldots, Y^{(D)} \leq y^{(D)}_{j_D}|x\right) = F(\alpha^{(1)}_{j_1} - x_1\beta_1, \ldots, \alpha^{(D)}_{j_D} - x_D\beta_D)$$ 
but then the question becomes of whether the data provides enough joint variation in indices $(x_1\beta_1, \ldots, x_D\beta_D)$ to identify $F$ on the whole support $\mathcal{E}$ of $\varepsilon$. 
The issue is that some (potentially each) $x_{d}$ could share all its covariates with another process.  In this case $(\alpha^{(1)}_{j_1} - x_1\beta_1,  \ldots, \alpha^{(D)}_{j_D} - x_D\beta_D)'$ could take values only in a proper subset of $\mathcal{E}$ and could vary only in certain directions as we vary the values of covariates. Since at this identification stage $(\alpha^{(1)}_{j_1} - x_1\beta_1, \ldots, \alpha^{(D)}_{j_D} - x_D\beta_D)'$ is observed, one could try and assess whether this vector covers the whole support $\mathcal{E}$. What we do is present conditions under which this is guaranteed. The illustration of our idea is given in Figure \ref{idCDF} for $D=2$. In one dimension (e.g. for $\varepsilon_1$) we ensure that $\alpha^{(1)}_{j_1} - x_1\beta_1$ can cover the whole marginal support of $\varepsilon_1$ (can be checked using conditions of Theorem \ref{corollary:Fd}). In the other dimension (e.g. for $\varepsilon_2$) we can require an exclusive covariate with non-zero coefficient -- without a loss of generality $x_{2,1}$ -- that can provide enough own variation in $\alpha^{(2)}_{j_2} - x_2\beta_2$ while keeping $x_1\beta_1$ fixed. In Figure \ref{idCDF} this variation is shown using vertical arrows. Once $x_1\beta_1$ is fixed, this variation can be checked to cover both lower and upper boundaries of $\mathcal{E}$ (either finite or infinite) by checking whether 
$\sup_{j_2} \sup_{x_{2,1}} F(\alpha^{(1)}_{j_1} - x_1\beta_1, \alpha^{(2)}_{j_2} - x_2\beta_2)$ coincides with $F(\alpha^{(1)}_{j_1} - x_1\beta_1)$ (upper boundary) and whether $\inf_{j_2} \inf_{x_{2,1}} F(\alpha^{(1)}_{j_1} - x_1\beta_1, \alpha^{(2)}_{j_2} - x_2\beta_2)$ is 0 (lower boundary). For general $D$, this identification strategy  can be translated into the requirements on exclusive covariates in $D-1$ processes.



\begin{theorem} \label{th:semiparamident3} Suppose all conditions of Theorem \ref{corollary:Fd} hold for each $d=1,\ldots,D$ and, hence, all the index parameters (subject to normalizations), thresholds, marginal c.d.f.s are identified.

In addition, suppose that 
\begin{itemize}
    \item[(a)] $\varepsilon$ is independent of $x$;
    
    \item[(b)]     at least $D-1$ processes -- without loss of generality processes $2$ to $D$ -- have $x_{d,1}$, $d=2,\ldots, D$, as an exclusive covariate with support large enough\footnote{{It does not have to be infinite -- it depends on the support of the underlying $\varepsilon$.}} to ensure that for some $(j_1,j_2, \ldots, j_D)$, for each $m=2,\ldots,D$,  
   \begin{equation}
   \label{joint1}
   \inf_{x_{m,1} \mid (x_{k})_{k=1}^{m-1}, x_{m,-1}} P\left(\cap_{k=1}^{m} (Y^{(k)} \leq y^{(k)}_{j_k}) \mid (x_{k})_{k=1}^{m-1}, x_{m} \right) = 0
   \end{equation}
    \begin{multline}
    \label{joint2}
    \sup_{x_{m,1} \mid (x_{k})_{k=1}^{m-1}, x_{m,-1}} P\left(\cap_{k=1}^{m} (Y^{(k)} \leq y^{(k)}_{j_k}) \mid (x_{k})_{k=1}^{m-1}, x_{m} \right) \\
    = P\left(\cap_{k=1}^{m-1} (Y^{(k)} \leq y^{(k)}_{j_k}) \mid (x_{k})_{k=1}^{m-1} \right)  
    \end{multline}
    for any $(x_{k})_{k=1}^{m-1}$ such that $P\left(\cap_{k=1}^{m-1} (Y^{(k)} \leq y^{(k)}_{j_k}) \mid (x_{k})_{k=1}^{m-1} \right) \in (0,1)$.
    \end{itemize}

\end{theorem}

Conditions (\ref{joint1}) and (\ref{joint2}) guarantee that $(\alpha^{(1)}_{j_1} - x_1\beta_1, \ldots, \alpha^{(D)}_{j_D} - x_D\beta_D)'$ for some $j_1,\ldots,j_D$ when taken in any direction $\lambda$ in $\mathbb{R}^D$ can reach the boundary of $\mathcal{E}$ in both positive and negative directions of $\lambda$.

To illustrate the progressive restrictiveness of the identification conditions outlined in Theorems  \ref{th:semiparamident3} through  \ref{th:semiparamident3}, we construct four nested data-generating processes (DGPs) for a bivariate ($D=2$) lattice model, each building sequentially on its predecessor. Each latent process contains a two-dimensional covariate vector associated with  $\beta_1 = \beta_2 = (1, 0.5)'$. In each dimension,  there are three ordered responses   and the threshold differences are 2,  Suppose the vector of unobservables is independent of covarioates and has a joint normal distribution.

In \textit{DGP 1}, covariates are defined as $x_1 = x_2=(x_{common1}, x_{common2})$, where $x_{common1}, x_{common2} \sim \text{Uniform}[-0.5, 0.5]$ and $x_{common1}, x_{common2}$ are not perfectly linearly related. This DGP provides limited support for $x_1\beta_1$ and $x_2\beta_2$ (it is within $[-0.75,0.75]$). Theorem 1 is satisfied, which ensures identification of $\beta_{1}$, $\beta_{2}$ up to scale, but fails to meet the conditions of Theorem \ref{th:semiparamident2} as it lacks overlaps in choice probabilities for threshold differences, Indeed, $P(Y^{c_d} \leq y^{(d)}_1|x_d) \in [\alpha^{(d)}_1-0.75,\alpha^{(d)}_1+0.75]$  whereas $P(Y^{c_d} \leq y^{(d)}_2|x_d) \in [2+\alpha^{(d)}_1-0.75,2+\alpha^{(d)}_1+0.75]=[\alpha^{(d)}_1
+1.25,\alpha^{(d)}_1+2.75]$ with $[\alpha^{(d)}_1-0.75,\alpha^{(d)}_1+0.75]$ and $[\alpha^{(d)}_1
+1.25,\alpha^{(d)}_1+2.75]$ obviously not overlapping. The narrow range of the indices precludes the probability matching required by Theorem \ref{th:semiparamident2}.

\textit{DGP 2} extends the first by widening the support of covariates: $x_{common1} \sim \text{Uniform}[-2, 2]$,  $x_{common2} \sim \text{Uniform}[-0.5, 0.5]$ enabling overlaps in conditional probabilities (e.g., $P(Y^{c_d} \leq y^{(d)}_1|x_d) \in [\alpha^{(d)}_1-2.25,\alpha^{(d)}_1+2.25]$ and $P(Y^{c_d} \leq y^{(d)}_2|x_d) \in [\alpha^{(d)}_1-0.25,\alpha^{(d)}_1+4.25]$). This satisfies the conditions up to Theorem 2 but falls short of Theorem \ref{corollary:Fd}, as the support, while sufficient for probability matching, does not fully cover the interval (0, 1). The added restrictiveness stems from the need for broader support to align probabilities, yet the coverage remains incomplete.

\textit{DGP 3} further extends the second by setting $x_{common1} \sim \text{Laplace}$,  $x_{common2} \sim \text{Uniform}[-0.5, 0.5]$  to ensure full probability coverage over (0, 1), and incorporates a normalization $F_d(0) = 0.5$. This setup satisfies the conditions up to Theorem \ref{corollary:Fd} but fails Theorem \ref{th:semiparamident3}. as the absence of exclusive covariates prevents independent shifting of dimensions to capture joint dependence.

\textit{DGP 4} builds on the third by defining  $x_{excl1} \sim Laplace$, $x_{excl2} \sim Laplace$, $x_{common2} \sim \text{Uniform}[-0.5, 0.5]$ (the suppose of the distribution of $(x_{excl1},x_{excl2},x_{common2})$ has an interior in $\mathbf{R}^3$,  This allows independent shifting of dimensions 1 abd 2, satisfying the requirement of  Theorem \ref{th:semiparamident3} (note this theorem only requires independent shifting of one dimension but for simplicity we allow that in both dimensions). All the parameters including the joint c.d.f.  can then be identified  fully.

\subsection{Estimation}

In what follows, we briefly outline some possibilities for estimating parameters in semiparametric models. A theme of this section is to outline  existing univariate ordered response estimation methods  that generalize to lattice models.

\paragraph*{Two-step approach}

The idea of this method is to (i) first use existing estimation approaches for semiparametric univariate ordered response models to estimate index and threshold parameters at a suitable rate (albeit suboptimally as the dependence of the latent processes is ignored), and (ii) second, construct estimates of the joint c.d.f. using some well known statistical methods. 

We start by discussing which estimation approaches in the literature can be utilized in the first step. 

\cite*{lewbel2000} 
develops a semiparametric estimator for qualitative response models (binary, ordered, multinomial) allowing for unknown heteroscedasticity in the latent errors with respect to regressors, or instrumental variables for endogeneity. The method relies on a ``special regressor'' $v$ that is conditionally independent of the error $\varepsilon$ given other regressors $x$ (i.e., $F_{\varepsilon|v,x}(\varepsilon \mid v, x) = F_{\varepsilon|x}(\varepsilon \mid x)$), with large support. The estimator resembles OLS or 2SLS on a transformed response $y^* = [y - I(v < 0)] / f(v \mid x)$, where $f$ is the conditional density of $v$ given $x$, yielding for ordered response models $\sqrt{n}$-consistent and asymptotically normal estimates for coefficients $\beta$ and thresholds (for ordered models). 

To generalize \cite*{lewbel2000} to lattice models, we need to have $x_d=(v_d,w_d)$ with a continuous special regression $v_d$ with large support per dimension $d$ -- this would effectively extend our Assumption ... (and accommodating heteroscedasticty by allowing $\text{Var}(\varepsilon_d \mid x_d)$ to be arbitrary). 
The estimator would proceed marginally per dimension using \cite*{lewbel2000} ordered method to recover $\beta_d$ and thresholds $\alpha^{(d)}_j$. \cite*{lewbel2000} consider a univariate ordered response, hence the question of joint c.d.f. does not arise (note, however, that for multinomial choice the estimation of joint c.d.f. of unobservables is relevant but \cite*{lewbel2000} does not address it).

 \cite*{klein2002} approach analyzes the univariate model, estimates the index parameter in the first stage using kernel density estimates of the conditional probability of choosing below a certain level. In the second stage, the approach estimates threshold parameters using shift restrictions. We can extend this approach to multivariate lattice models because  the functional independence  of thresholds across dimensions allows us to apply stages 1 and 2 
marginally for each $d=1,\dots,D$, using univariate techniques and our  Assumption \ref{assn:ind} which mirrors a cre assumption of independence in \cite*{klein2002} and  leads to  $P(Y_c^d \leq y^{(d)}_j | x_d) = F_d(\alpha^{(d)}_j - x_d \beta_d)$. The estimators or index and threshold parameters obtained from this stage are $\sqrt{n}$-consistent and asymptotically normal.  

\cite*{chenkhan2003} derives rates of convergence for estimating index parameters in heteroskedastic discrete response models, assuming multiplicative heteroskedasticity $\varepsilon_i = \sigma(x_i) \cdot u_i$, where $u_i$ is homoskedastic and independent of $x_i$. For ordered response models with at least three categories, $\sqrt{n}$-consistent estimators are possible. To generalize \cite*{chenkhan2003} to lattice models, we can consider each dimesn ion $d$ separately and consider at least three responses in that dimension. at the same time, we can  generalize it to  multiplicative heteroskedasticity per dimension: $\varepsilon_d = \sigma_d(x_d) \cdot u_d$, where $u_d$ is homoskedastic and independent of $x_d$. The \cite*{chenkhan2003} estimator for index parameters and thresholds proceeds marginally per dimension. Marginal stages inherit rates from  \cite*{chenkhan2003}:  $\sqrt{n}$-consistent $\hat{\beta}_d$, $\hat{\alpha}^{(d)}_j$ for $M_d \geq 3$. 

 \cite*{Liu2024} proposes two simple semiparametric estimators for univariate ordered response models with an unknown error distribution $F_0$, achieving $\sqrt{n}$-consistent and asymptotically normal estimators of the index parameters  and thresholds. The first method (binary choice-based) constructs nonparametric maximum likelihood estimates (NPMLE) of $F_0$ from recast binary data, then uses moment conditions index and threshold parameters. The second method (full ordered data) extends this by incorporating all outcomes via a weighted NPMLE. Both enforce monotonicity of $F_0$ and use bootstrap for inference. In lattice models, one can  apply \cite*{Liu2024} methods marginally per dimension to estimate $\beta_d$ and thresholds $\alpha^{(d)}_j$ (up to scale/location). All these estimators will be $\sqrt{n}$-consistent and asymptotically normal.

Thus, all these approaches are suitable when one's goal is to estimate  index and thresholds parameters. Given these estimates, one can now proceed with the estimation of the joint c.d.f. $F$ in  the second stage (this, of course, is not addressed in the papers mentioned above due to the univariate nature of the problem there). Let us now discuss some specific approaches that can be used to obtain  $\widehat{F}$. 

One possible approach is the \textit{grid inversion} method that discretizes the error space and solves a constrained optimization problem. It is a direct, computationally intensive non-parametric method. Let us outline it for $D=2$.  Its idea is based on the fact that given $x_i=(x_{i1},x_{i2})$ and $(Y^{(c_1)}_{j_1}=y^{(1)}_{j_1}, Y^{(c_2)}_{j_2}=y^{(2)}_{j_2})$, 
the latent pair $(\varepsilon_{1i}, \varepsilon_{2i})$ lies in the rectangle $R_i=
\times_{d=1}^2 \left( \alpha^{(d)}_{j_d-1} - x_{id}\beta_d, \, \alpha^{(d)}_{j_d} - x_{id}\beta_d \right]$ and, hence, due to independence of errors from covariates, 
\begin{equation}
\label{bivariate}
 P(Y^{(c_1)}=y^{(1)}_{j_1}, Y^{(c_2)}=y^{(2)}_{j_2} \mid X=x) = \sum_{\ell_1=0}^1\sum_{\ell_2=0}^1 (-1)^{\ell_1+\ell_2} F(\alpha^{(1)}_{j_1 -\ell_1} - x_{i1}\beta_1,\alpha^{(2)}_{j_2-\ell_2} - x_{i2}\beta_2).
\end{equation}

In the sample each observation $i$ implies a rectangular  interval $
\widehat{R}_i=\widehat{\mathbf{\varepsilon}}_i \in  \times_{d=1}^2 \left( \hat{\alpha}^{(d)}_{j_{d}(i)-1} - x_{di}\widehat{\beta}_d, \hat{\alpha}^{(d)}_{j_{d} (i)} - x_{di}\widehat{\beta}_d \right]
$ for the residual $\widehat{\mathbf{\varepsilon}}_i$, where $j_{d}(i)$ is the observed category in dimension $d$ for $i$ (with $-\infty, +\infty$ boundaries). Let $\mathcal{G} = \{ (e_{1,k}, e_{2,\ell}) : 
k=1,\dots,K_1,\ \ell=1,\dots,K_2 \}$ 
be the set of unique lower/upper bounds from all such implied sample rectangles.  Let
$
\phi = \big( F(e_{1,k}, e_{2,\ell}) \big)_{k,\ell} 
\in \mathbb{R}^{K_1 K_2}
$
collect the unknown c.d.f. values on this grid. We want to find the probability mass assigned to each grid point such that the implied probabilities for each cell 
 match the empirical probabilities in the data as closely as possible.
To do this, for each distinct covariate pattern $x_g$ (group), define the empirical cell probabilities
\[
\widehat{\pi}_{j_1 j_2}(x_g) \equiv \widehat{P}(Y^{(c_1)}_{i}=y^{(1)}_{j_1(i)}, Y^{(c_2)}_{i}=y^{(2)}_{j_2(i)} \mid X=x_i)
= \frac{ 
  \sum_{i : x_i = x_g} 
  \mathbf{1}\{ Y^{(1)}_{i} = y^{(1)}_{j_1}, Y^{(2)}_{i} = y^{(2)}_{j_2} \}}{ \sum_{i }: 1(x_i = x_g) }.
\]

Then for each $(j_1,j_2,g)$,
\[
\widehat{\pi}_{j_1 j_2}(x_g) 
= \sum_{k,\ell} A_{j_1 j_2,g}(k,\ell)\, \phi_{k\ell} + u_{j_1 j_2,g},
\]
where $A_{j_1 j_2,g}(k,\ell) \in \{-1,0,1\}$ encodes which c.d.f. corner terms 
enter each rectangle probability using (\ref{bivariate}). Stacking over all $(j_1,j_2,g)$ yields
$A \phi = \widehat{\pi} + u,
$
where $\widehat{\pi}$ collects all empirical cell probabilities. We can estimate $\phi$ by solving $\widehat{\phi} 
= \arg\min_{\phi \in \mathbf{\Phi}} \|A \phi - \widehat{\pi}\|^2$, where the feasible set $\mathbf{\Phi}$ enforces the defining properties of a c.d.f.: 
\[
\mathbf{\Phi} = 
\left\{
\phi : 0 \le \phi_{k\ell} \le 1, \,
\phi_{k\ell} \text{ nondecreasing in } k \text{ and in } \ell
\right\}.
\]
Optionally we can include a smoothness penalty and optimize $\min_{\phi \in \mathbf{\Phi}}
\|A \phi - \widehat{\pi}\|^2 + \lambda \|D \phi\|^2$,
where $D$ is a finite-difference matrix. The estimator provides
\[
\widehat{F}(e_{1,k}, e_{2,\ell}) = \widehat{\phi}_{k\ell},
\]
which can be extended to a continuous surface by bilinear interpolation.

There are some variations of this method. E.g., instead of the grid determined by the implied rectangular regions, one can consider a completely exogenous sample-free grid. 


Another possible approach is the kernel smoothing approach. Just like the inversion grid method it uses the fact that  
 $\mathbf{\varepsilon}_{i}\in R_i$ given $x_i=(x_{i1},x_{i2})$ and $(Y^{(c_1)}_{j_1}=y^{(1)}_{j_1}, Y^{(c_2)}_{j_2}=y^{(2)}_{j_2})$, and with  with $R_i$ defined in the same way as in the grid inversion method.  In the sample each observation $i$ implies  $
\widehat{\mathbf{\varepsilon}}_i \in  \widehat{R}_i
$.  We can implement a simulated kernel density estimator, where for each observation 
$i$ we draw $S$ 
 random samples $(\widetilde{\varepsilon}^{(s)}_1,\widetilde{\varepsilon}^{(s)}_2)$
 uniformly from its rectangle $\widehat{R}_i$. We then pool all these $N \times S$ simulated points together. We then perform a standard bivariate kernel density estimation on this large pooled sample. The resulting density is an estimate of $f(\varepsilon_1,\varepsilon_2)$. 
We can then integrate this estimated density numerically to get the estimated c.d.f. .

Other possible approaches include nonparametric sieve estimator subject to suitable choice of base (for monotonicity-preserving properties) and nonparametric maximum likelihood estimator. We have implemented the grid inversion and the simulated kernel density estimator in simulations biut not the other approaches. 

\paragraph*{One-step approach}

If one is interested in estimating the joint c.d.f of unobservable $\varepsilon$ (for purposes of analysing policy intervention or other counterfactuals), then one could extend \cite*{Coppejans2007} originally developed for univariate ordered response models under independence of the error and covariates. In what follows, we extend it to multivariate ordered response models, describing the bivariate case for illustrational simplicity. Suppose we have a random sample $\left\{(y^{(1)(i)}, y^{(2)(i)},x_1^{(i)},x_2^{(i)}) \right\}_{i=1}^N$. The idea is to maximize the log-likelihood function
\begin{equation}
    \mathcal{L}(\theta) = \frac{1}{N} \sum_{i=1}^{N}\sum_{j_{1}=1}^{M_{1}}\sum_{j_{2}=1}^{M_{2}}1\left[(y^{(1)(i)},y^{(2)(i)})=(y_{j_{1}}^{(1)},y_{j_{2}}^{(2)})\right]\log(\ell^{(i)}_{j_{1},j_{2}}),  \; \; \text{ where } \label{eq:mle1}
\end{equation}
\begin{eqnarray}
\ell^{(i)}_{j_{1},j_{2}} &=& F\left(a_{j_{1}}^{(1)}-x^{(i)}_{1}b_{1},a_{j_{2}}^{(2)}-x^{(i)}_{2}b_{2} \right) - F\left(a_{j_{1}-1}^{(1)}-x^{(i)}_{1}b_{1},a_{j_{2}}^{(2)}-x^{(i)}_{2}b_{2}\right) \nonumber \\
&-& F\left(a_{j_{1}}^{(1)}-x^{(i)}_{1}b_{1},a_{j_{2}-1}^{(2)}-x^{(i)}_{2}b_{2} \right) + F\left(a_{j_{1}-1}^{(1)}-x^{(i)}_{1}b_{1},a_{j_{2}-1}^{(2)}-x^{(i)}_{2}b_{2} \right), \label{eq:mle2}
\end{eqnarray} for joint c.d.f. of unobservables $F$. \cite{Coppejans2007} uses a quadratic B-spline to estimate the c.d.f of unobservables. The multivariate analogy is tensor-product B-splines. For instance, in the bivariate case the tensor-product basis  consists of $S_1\cdot S_2$ products of polynomials $\mathcal{R}$ in the form 
\begin{equation*}
\mathcal{R}_{1;s_{1},S_{1}}(e_1;q_{1})\mathcal{R}%
_{2;s_{2},S_{2}}(e_2;q_{2}),\quad s_{1}=1,\ldots
,S_{1},\;s_{2}=1,\ldots ,S_{2},
\end{equation*} 
here calculated for specific values of $e_1$ and $e_2$, with $q_d$ denoting the degree of B-spline in dimension $d=1,2$. A general tensor-product B-spline, which approximates $F(e_1,e_2)$, is a linear combination of these base tensor-product
polynomials with coefficients $\{h_{s_{1}s_{2}}\}$, $s_{d}=1,\ldots ,S_{d}$, 
$d=1,2$:
\begin{equation*}
\sum_{s_{1}=1}^{S_{1}}%
\sum_{s_{2}=1}^{S_{2}}h_{s_{1}s_{2}}\mathcal{R}%
_{1;s_{1},S_{1}}(e_1;q_{1})\mathcal{R}_{2;s_{2},S_{2}}(e_2;q_{2}).
\end{equation*}
The linear constraints 
\begin{align*} h_{s_{1}s_{2}} & \leq h_{s_{1}+1,s_{2}}, \quad \forall \: s_1=1,\ldots, S_1-1, \quad \, s_2 =1,\dots,S_{2}\\
     h_{s_{1}s_{2}} & \leq h_{s_{1},s_{2}+1}, \quad \forall \: s_2=1,\ldots, S_2-1, \quad  \, s_1=1,\dots,S_{1}
\end{align*}    
guarantee monotonicity of the tensor-product B-spline in each dimension. Additionally, the linear constraints 
$$ 0 \leq h_{s_{1},s_{2}} \leq 1, \quad  \forall \, s_1, s_2
$$ 
guarantee natural c.d.f. bounds of 0 and 1.\footnote{For more details on shape constraints in tensor-product B-splines, see \cite*{bk2022}.} Linear equality constraints on $h_{s_1s_2}$ can impose normalization restrictions on $F_d$: ...

\section{Parametric specification}
\label{sec:param}

In practice a researcher may choose a parametric family to model the distribution of unobservables conditional on covariates. On one hand, choosing a parametric family may allow researcher to explicitly model the distribution of observables as that depending on $x$ and be able to identify all the primitives given the assumed (potentially complicated) dependence structure. The exact identification strategy and assumption behind it will depend on the assumed structure. On the other hand, a researcher may still opt for independent errors and covariates and rely on less stringent data requirements for identification that those given in Section \ref{sec:semiparam} and a simpler estimation approach. 

We illustrate the latter case focusing on the lattice ordered probit (Gaussian errors) case. 

\begin{assumption}[Joint normal errors]\label{assn:normal}
The vector $\varepsilon$ is independent of $x$ and follows $N(0,\Sigma)$ where $\Sigma$ has ones on the diagonal and correlation $\rho_{kl}$ for as an off-diagonal $(k,l)$-element.\footnote{Note we have already normalized the means and variances of $\varepsilon_d$, $d=1,\ldots, D$, as it is easy to show that otherwise that the  best hope is identification up to a scale and a shift.  Theses are also   usual scale/location normalizations used e.g. in multinomial  probit.)}
\end{assumption}

\subsection{Identification}

As expected, due to our ability to view decisions rules across different dimensions index and threshold parameters can be identified using the same rank condition commonly employed  in univariate ordered probit models. This is formally presented in Theorem \ref{th:probit_index} below. Its proof is well known and we replicate it in the Appendix purely for completeness. 

\begin{theorem}  [index parameters and thresholds]\label{th:probit_index}
Suppose Assumption \ref{assn:normal} holds. If for a fixed dimension $d$ there exist $k_d+1$ points $\{x_d^{(i)}\}_{i=1}^{k_d+1}\subset \mathcal{X}_d$ such that the matrix
\[
\begin{pmatrix}
1 & x_d^{(1)}\\
1 & x_d^{(2)}\\
\vdots & \vdots\\
1 & x_d^{(k_d+1)}
\end{pmatrix}
\]
has rank $k_d+1$, then $\beta_d$ and the thresholds $\{\alpha^{(d)}_j\}_{j=1}^{M_d-1}$ are identified.
\end{theorem}

Identification of  correlation coefficients $\rho_{d_1,d_2}$ in the multivariate lattice setting does not follow from any readily available results in the literature. We can carry out this identification in the  pairwise manner under supplementary variation/exclusion conditions. They are collected in Theorem \ref{th:rho_id} below 

\begin{theorem}[Identification of pairwise correlations]\label{th:rho_id}
Suppose Assumption \ref{assn:normal} holds and Theorem \ref{th:probit_index}'s conditions hold for dimensions $d_1$ and $d_2$. Then the correlation $\rho_{d_1,d_2}$ is identified if at least one of the following holds:
\begin{itemize}
  \item[(a)] there exists  $x^*_{d_1}$ such that for some $j=1,\ldots, M_d-1$ it holds that $P(Y^{c_1}\leq y^{(d_1)}_{j_1} |x^*_{d_1})=0.5$;
  \item[(b)] 
  There are points $x, \widetilde{x}, x^{\diamond} \in \mathcal{X}$ such that for some $j_{1}=1,\ldots, M_{d_1}-1$, $j_{2}=1,\ldots, M_{d_2}-1$, 
  \begin{align*}(P(Y^{c_2}\leq y^{(d_2)}_{j_2}|x_{d_2})-0.5)(P(Y^{c_2}\leq y^{(d_2)}_{j_2}|\widetilde{x}_{d_2})-0.5)& >0, \\ 
  (P(Y^{c_2}\leq y^{(d_2)}_{j_2}|x_{d_2})-0.5)(P(Y^{c_2}\leq y^{(d_2)}_{j_2}|{x}^{\diamond}_{d_2})-0.5)&>0,\\
  (P(Y^{c_1}\leq y^{(d_1)}_{j_1}|x_{d_1})-0.5)(P(Y^{c_1}\leq y^{(d_1)}_{j_1}|\widetilde{x}_{d_1})-0.5)& <0.
  \end{align*}
  \item[(c)] there exists a subvector  in $x_{d_1}$ --  without a loss of generality suppose it is $x_{d_1, 1:L_{d_1}}$, $L_{d_1}\geq 1$, -- such that at least of the parameters in $\beta_{d_1,1:L_{d_1}}$ is not zero and  and $x_{d_1, 1:L_{d_1}}$ is excluded from $x_{d_2}$ -- that is, 
$$x_{d_1,\ell} \, | \, x_{d_2} \text{ has a non-degenerate distribution}, \quad l=1, \ldots, L_{d_1}.$$ 
Let $\mathcal{X}_{d_1 d_2}$ denote the projection of $\mathcal{X}$ onto the $(k_{d_1}+k_{d_2})$-dimensional space of covarites in dimensions $d_1$ and $d_2$ and 
suppose there are two different points in $\mathcal{X}_{d_1 d_2}$that differ only in the value of covariates in the subvector $x_{d_1,1:L_{d_1}}$ -- denote them as $(x_{d_1,1:L_{d_1}}^{(h)}, x_{d_1,L_{d_1}+1:k_{d_1}}, x_{d_2})$, $h=1,2$, -- such that for some indices $j_{1} \leq M_{d_1}-1$, $j_{2} \leq M_{d_2}-1$,  
\begin{multline*} 
P\left(Y^{(d_1)} \leq y^{(d_1)}_{j_1},  Y^{(d_2)} \leq y^{(d_2)}_{j_2} \, | \, x_{d_1,1:L_{d_1}}^{(1)}, x_{d_1,L_{d_1}+1:k_{d_1}}, x_{d_2} \right) 
\neq  \\ 
P\left(Y^{(d_1)} \leq y^{(d_1)}_{j_1},  Y^{(d_2)} \leq y^{(d_2)}_{j_2} \, | \, x_{d_1,1:L_{d_1}}^{(2)}, x_{d_1,L_{d_1}+1:k_{d_1}}, x_{d_2} \right).
\end{multline*}
\end{itemize}
\end{theorem}
Condition (a) requires a covariate configuration where the latent index in one dimension ($d_1$) is exactly at some  threshold. It creates a ``pivot'' where the error $\varepsilon_{d_1}$ is symmetrically distributed around zero, making joint probabilities with dimension 2 purely a function of $\rho_{d_1,d_2}$'s influence on $\varepsilon_{d_2}$. Condition (b) requires sign-flipping covariates. Namely, it assumes  covariate variation creating ``same-sign'' indices in one dimension (both above or both below the median threshold) but ``sign-flipping'' in the other. There are other ways to formulate related sufficient conditions in this spirit but we have opted to present this one. Condition (c) is an IV-style exclusion: a covariate (or subvector) affects dimension $d_1$'s outcome (via assocated nonzero $\beta_{d_1}$'s) but not dimension $d_2$'s directly (exclusion from $x_{d_2}$). By having a variable that affects only one outcome, we can trace out how joint probabilities shift when one margin's latent index moves while the other stays fixed. This variation rotates the joint probability surface (enough to do it once), letting us solve for $\rho_{d_1,d_2}$.

\subsection{Estimation in the parametric model} 
Estimation in the parametric model is standard via maximum likelihood. The log-likelihood function is equal to that in equations (\ref{eq:mle1}) and (\ref{eq:mle2}) with a specified cumulative distribution function $F$. We use bivariate normal as the natural example of $F$, as in Assumption \ref{assn:normal}, so that $\boldsymbol{\varepsilon}=(\varepsilon_1,\varepsilon_2)'$ is jointly normal with mean \((0,0)'\), unit variances, and correlation \(\rho\). Given a random sample $\left\{(y^{(1)(i)}, y^{(2)(i)},x_1^{(i)},x_2^{(i)}) \right\}_{i=1}^N$ and  collecting  $\beta_1,\beta_2,\rho$ and all the thresholds in $\alpha$ in one parameter vector $\theta$, we can construct the  log-likelihood function 
\begin{eqnarray*}
\mathcal{L}(\theta) &=& \frac{1}{N} \sum_{i=1}^{N}\sum_{j_{1}=1}^{M_{1}}\sum_{j_{2}=1}^{M_{2}} 1\left[(y^{(1)(i)},y^{(2)(i)})=(y_{j_{1}}^{(1)},y_{j_{2}}^{(2)})\right]\log(\ell^{(i)}_{j_{1},j_{2}}(\theta)) = \frac{1}{N}\sum_{i=1}^{N}\log(\ell^{(i)}(\theta)), 
\end{eqnarray*} 
$$\text{ with } \quad \ell^{(i)}_{j_{1},j_{2}} = 
\sum_{t_1=0}^{1} \sum_{t_2=0}^{1} (-1)^{t_1 + t_2} \Phi_{2}\left( \alpha_{j_1 - t_1, j_2}^{(1)} - x_1^{(i)}\beta_1, \alpha_{j_1, j_2 - t_2}^{(2)} - x_2^{(i)}\beta_2; \rho \right)$$ where $\Phi(\cdot,\cdot;\rho)$ denotes the standard bivariate normal c.d.f. with correlation parameter $\rho$.

The maximum likelihood estimator (MLE) $\hat{\theta}$ solves the optimization problem $\max_{\theta} \mathcal{L}(\theta)$.\footnote{One can impose inequality constraints on $\alpha$ and $\rho$ and maximize a constrained likelihood, or, more straightforwardly, re-parameterize the likelihood to estimate ($\alpha^{(j)}_{1},\sqrt{\alpha^{(j)}_{2}-\alpha^{(j)}_{1}},\dots,\sqrt{\alpha_{M_{d}}^{(j)}-\alpha_{M_{d}-1}^{(d)}})$ and $\tanh{^-1}(\rho)$ so that constraints are enforced automatically.} Under the typical MLE regularity conditions (e.g. \citep*{newey1994}), we have $\sqrt{N}(\hat{\theta}-\theta_{0}) \overset{d}{\longrightarrow} \mathcal{N}(0,V)$, $V = \mathbb{E}\left[\frac{\partial \log(\ell^{(i)}(\theta_0))}{\partial \theta}\frac{\partial \log(\ell^{(i)}(\theta_0))}{\partial \theta'}\right]$. The natural plug-in sample-analogue estimator of $V$ provides a consistent estimator for the variance-covariance matrix.

\section{Simulations}
\label{sec:montecarlo}

We consider a bivariate ordered response model with
\begin{align}
Y^{*c_1}_{i} &= x_{1i}{\beta}_1 + \epsilon_{1i}, \label{eq:sim1intro} \\
Y^{*c_2}_{i} &= x_{2i}{\beta}_2 + \epsilon_{2i}. \label{eq:sim2intro}
\end{align}

\subsection{Semiparametric model}

Here we focus on \textit{two-step approaches}. We take the index and threshold parameters to be known (in reality, they would have been estimated consistently at $\sqrt{n}$ rate) and just focus on the estimation of the joint c.d.f. given these parameters. This allows us to compare the performance of different second-stage approaches in their pure form without first-stage inference.\footnote{Moreover, first-stage estimation approaches come from already existing literature.} We take both $x_{1i}$ and $x_{2i}$ to be univariate with respective $\beta_1 = 0.8$, $\beta_2 = -0.5$. We adopt $3 \times 3$ categorical outcomes with the thresholds determining the decision structure given by $\alpha^{(1)}_0=-\infty$, $\alpha^{(1)}_1=-1$, $\alpha^{(1)}_2=1$, $\alpha^{(1)}_3=+\infty$ for dimension 1 and $\alpha^{(2)}_0=-\infty$, $\alpha^{(2)}_1=-0.8$, $\alpha^{(2)}_2=0.8$, $\alpha^{(2)}_3=+\infty$.  for dimension 2. We take ${x}_1 \sim N(0,1)$, ${x}_2 \sim 0.5N(0,1) + 0.3{x}_1$, and $\boldsymbol{\varepsilon}$ is bivariate normal with mean zero, unit variances and the correlation coefficient 0.6.  We draw $S=10$ points from the rectanbgle associated with observation $i$.  

We compare the performance of our estimators on an $80 \times 80$ evaluation grid over $[-2.5, 2.5]^2$:  We use  $G=6,400$ to denote the number of points in the evaluation grid and $g$ to denote a particular point on this grid. As criteria we use 
 Root Mean Square Error $\sqrt{\frac{1}{G}\sum_{g=1}^n (\hat{F}(g) - F(g))^2}$ (RMSE), Kolmogorov-Smirnov (KS) distance $\max_g |\hat{F}(g) - F(g)|$, Cramér-von Mises (CvM) distance $\frac{1}{G}\sum_{g=1}^G (\hat{F}(g) - F(g))^2$ and correlation $\text{corr}(\hat{{F}}, {F})$. 
 

Table~\ref{tab:main_results_sim1} presents simulation results comparing both approaches on the average of the four metrics in 400 simulations.

\begin{table}[h!]
\centering
\caption{Overall performance comparison}
\label{tab:main_results_sim1}
\begin{threeparttable}
\begin{tabular}{lcccccc}
\hline
\hline
\textbf{Method} & \textbf{RMSE} & \textbf{KS Distance} & \textbf{CvM Distance} & \textbf{Correlation} \\
\hline
Grid Inversion & 0.07232 & 0.192247 & 0.005244  & 0.991955 \\
\quad & (0.003677) & (0.011666) & (0.000534) & (0.000509) \\
Kernel Smoothing & 0.026877 & 0.073202 & 0.000729  & 0.996752 \\
\quad & (0.002509) & (0.007906) & (0.000138) & (0.000420) \\
\hline
\end{tabular}
\begin{tablenotes}
\small
\item Note: All metrics are calculated on evaluation grid with $80 \times 80 = 6,400$ points and then averaged across 400 simulations. Parantheses contain standard deviations across simulations. Lower values indicate better performance for all metrics except correlation.
\end{tablenotes}
\end{threeparttable}
\end{table}
On the basis of these results,  the kernel smoothing method dominates the grid inversion method on the four metrics. We do not pursue further with regard to how well these methods do with regard to various regions (central vs tail ones) or which method performs better with regard to some specific distributional characteristics such as entropy, or tail mass but one could of course pursue this type of simulation analysis as well. It may very well be the case that the grid inversion method may perform better in other criteria as it explicitly incorporates the ordinal response structure through the design matrix $A$ and its associated constrained least squares formulation provides a globally optimal solution for the discrete approximation. Moreover. this 

\begin{figure}[htbp]
  \centering
  \begin{minipage}[b]{0.45\textwidth}
    \centering
    \includegraphics[width=\textwidth]{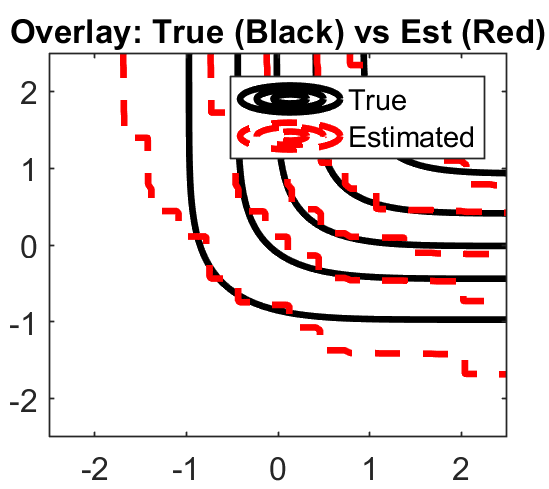}
    \caption{Grid inversion method}
    \label{fig:graph1}
  \end{minipage}
  \hfill
  \begin{minipage}[b]{0.45\textwidth}
    \centering
    \includegraphics[width=\textwidth]{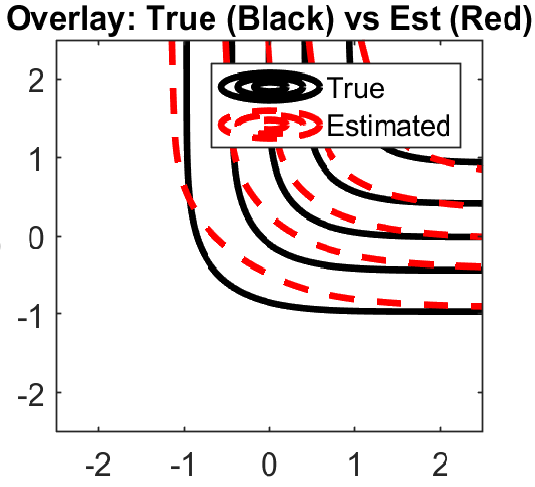}
    \caption{Kernel smoothing method}
    \label{fig:graph2}
  \end{minipage}
  \caption{Illustration of the joint c.d.f. estimation in step 2 of two-step approaches.The graphs present contour curves for the true and estimated c.d.f. }
\end{figure}

\subsection{Parametric model}

We now examine Monte Carlo simulations for the parametric case with normal errors. For the purposes of the simulations, we rewrite Equations (\ref{eq:sim1intro}) and (\ref{eq:sim2intro}) as $$Y_{i}^{*c_{1}} = x_{i}\beta_{1} + w_{1i}\gamma_{1} + \varepsilon_{1i}, \qquad Y_{i}^{*c_{2}} = x_{i}\beta_{2} + w_{2i}\gamma_{2} + \varepsilon_{2i}$$ to distinguish exclusive ($w$) and non-exclusive ($x$) covariates. We explore a first scenario with no exclusive covariates ($\gamma_{1}=\gamma_{2}=0$), a second scenario with an exclusive covariate in one latent process, and a third scenario with exclusive covariates in both latent processes. Each simulation design uses 400 independent random samples of size 1,000. A summary of the following results is that in all models, which vary in their number of discrete values $M$, type of regressors (discrete or continuous) and exclusivity of regressors, all parameters are estimated with essentially no bias; threshold and index parameters are estimated more precisely than the correlation parameter.

\subsubsection*{Parametric Design 1: 2$\times$2 structure, no excluded regressors} 

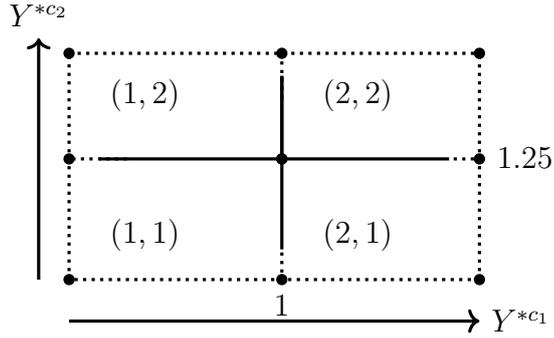
\begin{figure}[tbp]
\centering
\caption{Latent variable space in Parametric Design 1}
\begin{tikzpicture}[scale=0.2]
\draw [very thick, dotted ] (15.2,-4) -- (13.2,-4);
\draw [very thick, dotted ] (-8.2,-4) -- (-12,-4);
\draw [very thick, dotted ] (2,-10.2) -- (2,-12);
\draw [very thick, dotted ] (2,0) -- (2,3);
\draw [very thick, dotted ] (-12,-12) -- (-12,3);
\draw [very thick, dotted ] (15,-12) -- (15,3);
\draw [very thick, dotted ] (-12,3) -- (15,3);
\draw [very thick, dotted ] (15,-12) -- (-12,-12);

\draw [very thick ] (-10,-4) -- (13,-4);
\draw [very thick ] (2,-10) -- (2,1.5);
\draw [very thick ] (2,-4) -- (2,1.5);

\node at (-12,-4) [circle,fill,inner sep=1.5pt, ]{};
\node at (2,-4) [circle,fill,inner sep=1.5pt, ]{};
\node at (2,-4) [circle,fill,inner sep=1.5pt, ]{};
\node at (15,-4) [circle,fill,inner sep=1.5pt, ]{};

\node at (15,-12) [circle,fill,inner sep=1.5pt, ]{};
\node at (2,-12) [circle,fill,inner sep=1.5pt, ]{};
\node at (-12,-12) [circle,fill,inner sep=1.5pt, ]{};

\node at (-12,3) [circle,fill,inner sep=1.5pt, ]{};
\node at (15,3) [circle,fill,inner sep=1.5pt, ]{};
\node at (2,3) [circle,fill,inner sep=1.5pt, ]{};

\draw [, font = ] (-7,-9) node (00) {$(1,1)$};
\draw [, font = ] (7,-9) node (20) {$(2,1)$};
\draw [, font = , below] (-7,2) node (01) {$(1,2)$};
\draw [, font = , below] (7,2) node (21) {$(2,2)$};

\draw[very thick,-> ] (-12,-14.75) -- (15,-14.75) node[right] {$Y^{*c_1}$};
\draw[very thick,-> ] (-14,-12) -- (-14,4) node[above] {$Y^{*c_2}$};

\node [below, thick ] at (2,-12.25) { $1$};
\node [below, thick ] at (18,-2.5) { $1.25$};
\end{tikzpicture}
\label{fig:figSim3}
\end{figure}

We investigate parametric estimation without exclusive covariates by setting $\gamma_1=\gamma_2=0$, removing $w_1$ and $w_2$. We set $\beta_1=3$, $\beta_2=2.5$, $\rho=0.33$, and use a $2\times2$ non-lattice structure with thresholds $\alpha^{(1)}_{1}=1$ and $\alpha_{1}^{(2)}=1.25$ (see Figure~\ref{fig:figSim3}). The common regressor $x$ follows a uniform $[-4,4]$ distribution.

Table~\ref{tabSIM1new} Panel 1 reports mean and standard deviation of parameter estimates. The  method estimates all parameters with minimal bias.  Estimates of $\rho$ are less precise due to the absence of excluded regressors.

\begin{table}[tbp]
\caption{Parametric simulation results} \label{tabSIM1new}
\centering
\begin{threeparttable}
\begin{tabular}{cccccccccccc}
\toprule 
\textbf{Parameter} &&& \textbf{Truth} &&& \multicolumn{1}{c}{\textbf{Mean}} &&& \multicolumn{1}{c}{\textbf{SD}}\\
\hline \\
\textit{Panel 1:}\\
$\beta_{1}$                   &&& 3                     &&& 3.08&&& 0.35    \\
$\beta_{2}$                   &&& 2.5                   &&& 2.53&&& 0.23  \\
$\rho$                        &&& 0.33                  &&& 0.34&&& 0.14 \\
$\alpha_{1}^{(1)}$              &&& 1                 &&&  1.02&&& 0.16 \\
$\alpha_{1}^{(2)}$              &&& 1.25                &&&  1.26&&& 0.15  \\\\

\textit{Panel 2:}\\ 
$\beta_{1}$  &&& 2 &&& 2.02&&& 0.10 \\
$\gamma_{1}$ &&& -3  &&& -3.03&&& 0.15 \\
$\beta_{2}$  &&& 3   &&& 3.01&&& 0.16 \\
$\rho$       &&& 0.25 &&& 0.26&&& 0.09 \\
$\alpha^{(1)}_{1}$ &&& -1.5 &&& -1.51&&& 0.12 \\
$\alpha^{(1)}_{2}$ &&& 0.6 &&& 0.60&&& 0.10 \\
$\alpha^{(1)}_{3}$ &&& 4 &&& 4.04 &&&0.21 \\
$\alpha^{(2)}_{1}$ &&& -2.5 &&& -2.50&&& 0.15 \\
$\alpha^{(2)}_{1}$&&& 2 &&& 2.02 &&&0.13 \\\\

\textit{Panel 3}\\
$\beta_{1}$  &&& 1.75 &&&  1.75 &&&0.08  \\
$\gamma_{1}$ &&& -2.75  &&&  -2.76 &&& 0.12 \\
$\beta_{2}$  &&& 2.5   &&&  2.65 &&&0.39  \\
$\gamma_{2}$ &&& -4  &&&  -4.24 &&&0.64 \\
$\delta_{2}$ &&& 2   &&&  2.11 &&&0.32  \\
$\rho$       &&& 0.5 &&&  0.53 &&&0.19  \\
$\alpha^{(1)}_{1}$ &&& -7 &&& -7.04&&& 0.31 \\
$\alpha^{(1)}_{2}$ &&& -5 &&& -5.01 &&&0.21 \\
$\alpha^{(1)}_{3}$ &&& -0.75 &&& -0.75&&& 0.08 \\
$\alpha^{(1)}_{4}$ &&& 2.5 &&& 2.51&&& 0.13 \\
$\alpha^{(1)}_{5}$ &&& 4 &&& 4.00 &&&0.18 \\
$\alpha^{(2)}_{1}$ &&& -2 &&& -2.10 &&&0.33 \\
\bottomrule 
\end{tabular} \vspace*{-0.1cm}
\begin{tablenotes}[flushleft]
\footnotesize
\item Notes: Table \ref{tabSIM1new} reports the sample mean and sample standard deviations of the estimates of the parameters, over 400 samples. The panels correspond to the respective designs.
\end{tablenotes}
\end{threeparttable}
\end{table}

\subsubsection*{Parametric Design 2: 4$\times$3 with one excluded covariate}
\label{subsection:des2}

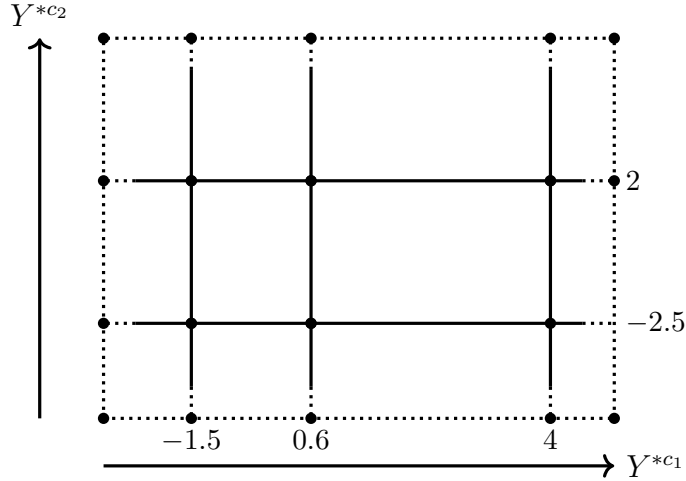
\begin{figure}[tbp]
\centering
\caption{Latent variable space for two equations: Design 2}
\begin{tikzpicture}[scale=0.42]

\draw [very thick] (-5,-3) -- (-3.25,-3);
\draw [very thick] (-3.25,-3) -- (8,-3);
\draw [very thick] (-5,1.5) -- (8,1.5);
\draw [very thick] (8,-3) -- (9,-3);
\draw [very thick] (8,1.5) -- (9,1.5);

\draw [very thick] (-3.25,-5) -- (-3.25,1.5);
\draw [very thick] (0.5,1.5) -- (0.5,5);
\draw [very thick] (0.5,-5) -- (0.5,-3);
\draw [very thick] (0.5,-3) -- (0.5,1.5);
\draw [very thick] (-3.25,1.5) -- (-3.25,5);
\draw [very thick] (8,-5) -- (8,5);

\draw [very thick, dotted ] (-3.25,-5) -- (-3.25,-6);
\draw [very thick, dotted ] (0.5,-5) -- (0.5,-6);
\draw [very thick, dotted ] (8,-5) -- (8,-6);
\draw [very thick, dotted ] (9,-3) -- (10,-3);
\draw [very thick, dotted ] (9,1.5) -- (10,1.5);
\draw [very thick, dotted ] (8,5) -- (8,6);
\draw [very thick, dotted ] (-3.25,5) -- (-3.25,6);
\draw [very thick, dotted ] (0.5,5) -- (0.5,6);

\draw [very thick, dotted ] (-5,-3) -- (-6,-3);
\draw [very thick, dotted ] (-5,1.5) -- (-6,1.5);

\node at (-3.25,-3) [circle,fill,inner sep=1.5pt]{};
\node at (-3.25,-3) [circle,fill,inner sep=1.5pt]{};
\node at (-3.25,1.5) [circle,fill,inner sep=1.5pt]{};
\node at (0.5,1.5) [circle,fill,inner sep=1.5pt]{};

\node at (0.5,-3) [circle,fill,inner sep=1.5pt]{};
\node at (0.5,-3) [circle,fill,inner sep=1.5pt]{};
\node at (0.5,1.5) [circle,fill,inner sep=1.5pt]{};
\node at (-3.25,1.5) [circle,fill,inner sep=1.5pt]{};

\node at (8,-3) [circle,fill,inner sep=1.5pt]{};
\node at (8,1.5) [circle,fill,inner sep=1.5pt]{};
\node at (8,1.5) [circle,fill,inner sep=1.5pt]{};

\node at (-3.25,-6) [circle,fill,inner sep=1.5pt]{};
\node at (0.5,-6) [circle,fill,inner sep=1.5pt]{};
\node at (8,-6) [circle,fill,inner sep=1.5pt]{};

\node at (-6,-6) [circle,fill,inner sep=1.5pt]{};
\node at (-6,-3) [circle,fill,inner sep=1.5pt]{};
\node at (-6,1.5) [circle,fill,inner sep=1.5pt]{};
\node at (-6,6) [circle,fill,inner sep=1.5pt]{};

\node at (0.5,6) [circle,fill,inner sep=1.5pt]{};
\node at (-3.25,6) [circle,fill,inner sep=1.5pt]{};
\node at (8,6) [circle,fill,inner sep=1.5pt]{};
\node at (10,6) [circle,fill,inner sep=1.5pt]{};
\node at (10,1.5) [circle,fill,inner sep=1.5pt]{};
\node at (10,-6) [circle,fill,inner sep=1.5pt]{};

\node [below, thick] at (-3.25,-6) {\small $-1.5$};
\node [below, thick] at (0.5,-6) {\small $  0.6$};
\node [below, thick] at (8,-6) {\small $ 4$};
\node [right, thick] at (10,-3) {\small $ -2.5$};
\node [right, thick] at (10,1.5) {\small $ 2$};

\draw[very thick,-> ] (-6,-7.5) -- (10,-7.5) node[right] {$Y^{*c_1}$};
\draw[very thick,-> ] (-8,-6) -- (-8,6) node[above] {$Y^{*c_2}$};

\draw [very thick, dotted ] (-6,-6) -- (-6,6);
\draw [very thick, dotted ] (-6,-6) -- (10,-6);
\draw [very thick, dotted ] (10,-6) -- (10,6);
\draw [very thick, dotted ] (10,6) -- (-6,6);

\end{tikzpicture}
\label{figSimDesign2}
\end{figure}

In the second simulation design, we extend the number of discrete values $M_{d}$ in both dimensions. The discrete dependent variable $Y^{c_1}$ can take four values and $Y^{c_2}$ can take three values. This generates a 4$\times$3 structure, illustrated in Figure \ref{figSimDesign2}. The common covariate $x$ follows a uniform $[-2,2]$ distribution (alternatively we could have taken it to be discrete). The covariate $w_{1}$ is a discrete random variable taking values -2.5, -1.5, -0.5 and 0.5 with equal probability. We set $\gamma_2=0$ thus effectively removing $w_{2}$ in the second equation. The parameter values are $\beta_{1} = 2, \gamma_{1} = -3, \beta_{2} = 3$ and $\rho = 0.25$. 

Table \ref{tabSIM1new} Panel 2 lists the across-simulation means and standard deviations of the index parameters, thresholds, and the correlation coefficient. The bivariate ordered probit method estimates all parameters with no bias. The correlation parameter remains the least precise estimate across parameters.

\subsubsection*{Parametric Design 3: 6$\times$2}
\label{subsection:des3}
\begin{figure}[t!]
\centering
\caption{Latent variable space for two equations: Design 3}
\begin{tikzpicture}[scale=0.4]

\draw[very thick,-> ] (-9,-11) -- (16,-11) node[right] {$Y^{*1}$};
\draw[very thick,-> ] (-12,-9) -- (-12,9) node[above] {$Y^{*2}$};


\node at (-10,-9) [circle,fill,inner sep=1.5pt]{};
\node at (-10,1) [circle,fill,inner sep=1.5pt]{};
\node at (-10,8) [circle,fill,inner sep=1.5pt]{};

\node at (-6,-9) [circle,fill,inner sep=1.5pt]{};
\node at (-6,1) [circle,fill,inner sep=1.5pt]{};
\node at (-6,1) [circle,fill,inner sep=1.5pt]{};
\node at (-6,8) [circle,fill,inner sep=1.5pt]{};

\node at (-2,-9) [circle,fill,inner sep=1.5pt]{};
\node at (-2,1) [circle,fill,inner sep=1.5pt]{};

\node at (-2,1) [circle,fill,inner sep=1.5pt]{};
\node at (-2,8) [circle,fill,inner sep=1.5pt]{};

\node at (2,-9) [circle,fill,inner sep=1.5pt]{};
\node at (2,1) [circle,fill,inner sep=1.5pt]{};
\node at (2,1) [circle,fill,inner sep=1.5pt]{};
\node at (2,8) [circle,fill,inner sep=1.5pt]{};

\node at (5,-9) [circle,fill,inner sep=1.5pt]{};
\node at (5,1) [circle,fill,inner sep=1.5pt]{};
\node at (5,1) [circle,fill,inner sep=1.5pt]{};
\node at (5,8) [circle,fill,inner sep=1.5pt]{};

\node at (9,-9) [circle,fill,inner sep=1.5pt]{};
\node at (9,1) [circle,fill,inner sep=1.5pt]{};
\node at (9,1) [circle,fill,inner sep=1.5pt]{};
\node at (9,8) [circle,fill,inner sep=1.5pt]{};

\node at (15,-9) [circle,fill,inner sep=1.5pt]{};
\node at (15,1) [circle,fill,inner sep=1.5pt]{};
\node at (15,8) [circle,fill,inner sep=1.5pt]{};

\draw [very thick, dotted ] (-10,-9) -- (-10,8);
\draw [very thick, dotted ] (-10,-9) -- (15,-9);
\draw [very thick, dotted ] (15,-9) -- (15,8);
\draw [very thick, dotted ] (-10,8) -- (15,8);

\draw [very thick] (-6,-7) -- (-6,6);
\draw [very thick] (-2,-7) -- (-2,-2);
\draw [very thick] (-2,-2) -- (-2,6);
\draw [very thick] (2,-7) -- (2,6);
\draw [very thick] (5,-7) -- (5,6);
\draw [very thick] (9,-7) -- (9,6);

\draw [very thick] (-8,1) -- (-6,1);
\draw [very thick] (-6,1) -- (2,1);
\draw [very thick] (2,1) -- (5,1);
\draw [very thick] (5,1) -- (9,1);
\draw [very thick] (9,1) -- (13,1);

\draw [very thick, dotted ] (-6,-9) -- (-6,-7);
\draw [very thick, dotted ] (-2,-9) -- (-2,-7);
\draw [very thick, dotted ] (2,-9) -- (2,-7);
\draw [very thick, dotted ] (5,-9) -- (5,-7);
\draw [very thick, dotted ] (9,-9) -- (9,-7);

\draw [very thick, dotted ] (-6,6) -- (-6,8);
\draw [very thick, dotted ] (-2,6) -- (-2,8);
\draw [very thick, dotted ] (2,6) -- (2,8);
\draw [very thick, dotted ] (5,6) -- (5,8);
\draw [very thick, dotted ] (9,6) -- (9,8);

\draw [very thick, dotted ] (-10,1) -- (-8,1);
\draw [very thick, dotted ] (13,1) -- (15,1);

\node [above, thick] at (-6,-11) {\large $-7$};
\node [above, thick] at (-2,-11) {\large $-5$};
\node [above, thick] at (2,-11) {\large $-0.75$};
\node [above, thick] at (5,-11) {\large $2.5$};
\node [above, thick] at (9,-11) {\large $4$};

\node [right, thick] at (15,1) {\large $-2$};

\end{tikzpicture}
\label{fig:des2}
\end{figure}
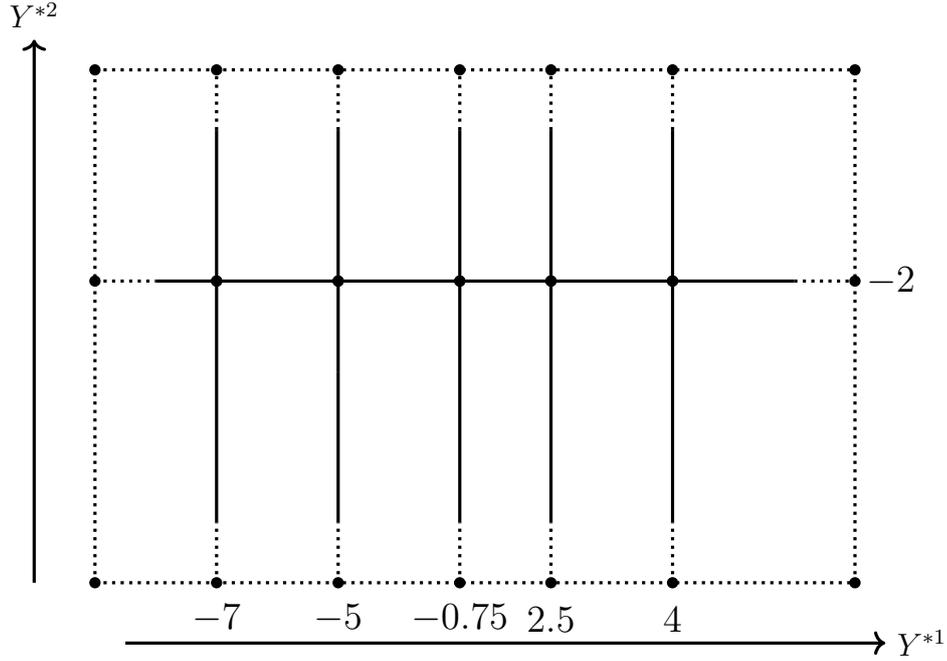

We consider a design that creates a 6$\times$2 structure on the latent variable space. Figure \ref{fig:des2} illustrates the threshold structure and the values of thresholds. In this design, the common regressor $x$ is drawn from uniform $[-2,2]$ and both latent equations have excluded regressors $w_{1},w_{2} \overset{iid}{\sim} t_{7}$. We also include an additional regressor $z_{2}$ in equation 2, drawn from a logistic (3,2) distribution. The parameter corresponding to $z_{2}$ is denoted $\delta_{2}$, so that the latent equations read \begin{eqnarray*}
Y^{*c_1}_{i} &=& x_i\beta_{1} + w_{1i}\gamma_{1} + \varepsilon_{1i} \\
Y^{*c_2}_{i} &=& x_i\beta_{2} + w_{2i}\gamma_{2} + \varepsilon_{2i} + z_{2i}\delta_{2}
\end{eqnarray*}

The parameter values are $\beta_1 = 1.75$, $\beta_2 = 2.5$, $\gamma_1 = -2.75$, $\gamma_2=-4$, and $\delta_{2} = 2$. Table \ref{tabSIM1new} presents the results for the index parameters, thresholds, and the correlation coefficient. Generally, all parameters are estimated with low bias, though slightly more bias in the index parameters in dimension two than one.

\section{Application} 
\label{sec:appl}

Now we study the main factors driving self-reported health and happiness as well co-movement in unobservables driving them. To do this, we pool data on the United States of America and Canada from six waves of the World Values Survey \citep{inglehart2014}. The results are also fully robust to the use of a different dataset: the National Health and Nutrition Examination Survey (NHANES). A full description of these two datasets and variable construction is provided in the Appendix.

The specification considered follows the standard setup described in the paper. Namely, for latent physical health ($p$) and sadness ($m$) variables $Y^{*}_{p}$ and $Y^{*}_{m}$ respectively, we have \begin{eqnarray*}
Y_{p}^{*} &=& x\beta_{p} + w_{p}\gamma_{p} + \varepsilon_{p} \\
Y_{m}^{*} &=& x\beta_{m} + w_{m}\gamma_{m} + \varepsilon_{m}
\end{eqnarray*}
with common row of covariates $x$ and exclusive covariates $w_{p}$ and $w_{m}$ in those two processes. The discrete dependent variable for health we use takes three values: 0, 1, and 2. The value 0 represents a self-reported ``State of health'' as ``fair'', ``poor'', or ``very poor''. The value 1 represents a report of ``good'', and the value 2 a report of ``very good''. The dependent variable for happiness again takes values 0, 1, and 2. In this case, a value of 0 represents a self-reported ``Feeling of happiness'' as ``very happy''. A value of 1 represents a reporting of ``quite happy'', and a value of 2 a reporting of either ``not very happy'' or ``not at all happy''. It is coded so that higher values reflect \textit{lower} self-reported happiness.

The common set of regressors $x$ includes the variabless: male, white, a college education dummy, age, regional dummies, and 5 dummies for income brackets. The 5 income brackets are: (1) less than \$20,00; (2) between \$20,000 and \$35,000; (3) between \$35,000 and \$50,000; (4) between \$50,000 and \$ 75,000; and (5) greater than \$100,000. We include no excluded health regressors, so that $w_{p} = 0$, but include dummies for employment status and living with a partner as excluded happiness regressors, $w_{m}$. 

The coefficients from the lattice bivariate ordered probit regression are provided in Table \ref{tabHEALTH1}. The signs of the regression coefficients are as expected. Positive partial effects on the probability of a better health (above a certain level) are given by varaiables that include dummies for white ethnicity and college education, and higher income brackets. The variables that have a positive impact on the probability of a higher happiness (or, equivalently, lower sadness) include living with a partner and  higher income brackets.  The effect of employment status on the probability of a higher happiness  appears negative but is not statistically significant. 

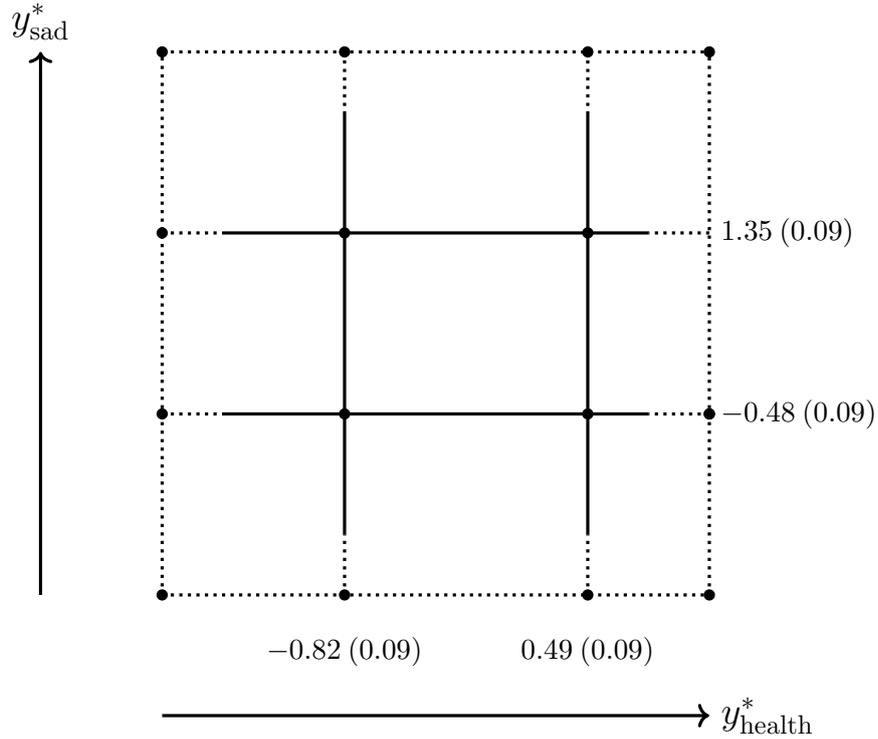
\begin{figure}
\centering
\begin{tikzpicture}[scale=0.4]


\draw [very thick,dotted] (-9,-9) -- (-9,9);
\draw [very thick,dotted] (9,-9) -- (9,9);
\draw [very thick,dotted] (-9,-9) -- (9,-9);
\draw [very thick,dotted] (-9,9) -- (9,9);

\draw [very thick,dotted] (-9,-3) -- (-7,-3);
\draw [very thick,dotted] (-9,3) -- (-7,3);
\draw [very thick,dotted] (9,-3) -- (7,-3);
\draw [very thick,dotted] (9,3) -- (7,3);

\draw [very thick,dotted] (-3,-9) -- (-3,-7);
\draw [very thick,dotted] (5,-9) -- (5,-7);
\draw [very thick,dotted] (-3,9) -- (-3,7);
\draw [very thick,dotted] (5,9) -- (5,7);

\draw [very thick] (-7,-3) -- (7,-3);
\draw [very thick] (-7,3) -- (7,3);

\draw [very thick] (-3,-7) -- (-3,7);
\draw [very thick] (5,-7) -- (5,7);

\node [right, thick, ] at (9,-3) {\small $-0.48 \: (0.09)$};
\node [right, thick,] at (9,3) {\small $1.35 \:(0.09)$};
\node [below, thick,] at   (-3,-10) {\small $-0.82 \: (0.09)$};
\node [below, thick,] at (5,-10) {\small $0.49 \: (0.09)$};

\node at (-9,-9) [circle,fill,inner sep=1.5pt]{};
\node at (-9,-3) [circle,fill,inner sep=1.5pt]{};
\node at (-9,3) [circle,fill,inner sep=1.5pt]{};
\node at (-9,9) [circle,fill,inner sep=1.5pt]{};

\node at (-3,-9) [circle,fill,inner sep=1.5pt]{};
\node at (-3,-3) [circle,fill,inner sep=1.5pt]{};
\node at (-3,3) [circle,fill,inner sep=1.5pt]{};
\node at (-3,9) [circle,fill,inner sep=1.5pt]{};

\node at (5,-9) [circle,fill,inner sep=1.5pt]{};
\node at (5,-3) [circle,fill,inner sep=1.5pt]{};
\node at (5,3) [circle,fill,inner sep=1.5pt]{};
\node at (5,9) [circle,fill,inner sep=1.5pt]{};

\node at (9,-9) [circle,fill,inner sep=1.5pt]{};
\node at (9,-3) [circle,fill,inner sep=1.5pt]{};
\node at (9,-3) [circle,fill,inner sep=1.5pt]{};
\node at (9,9) [circle,fill,inner sep=1.5pt]{};

\draw[very thick,-> ] (-9,-13) -- (9,-13) node[right] {\large $y_{\text{health}}^{*}$};
\draw[very thick,-> ] (-13,-9) -- (-13,9) node[above] {\large $y_{\text{sad}}^{*}$};

\end{tikzpicture}
\caption{\textit{Estimates from Lattice Bivariate Probit: Health and Happiness}}
\label{figHealthLatt}
\end{figure}

The estimated value of the correlation between the two model errors is negative, which is consistent with our expectations that shocks increasing health would tend to decrease sadness and shocks decreasing health would tend to increase sadness. The thresholds produced are shown in Figures \ref{figHealthLatt}. 

Detailed estimation results for index parameters are given in Table 
\begin{table}
\renewcommand{\arraystretch}{1}
\centering
\begin{threeparttable}
\caption{\textsc{Estimation Coefficients: Health and Happiness}}
\begin{tabular}{ccccc}
\hline
\hline 
\\
\textit{Dependent Variable: Health}\\
\hline
\textsc{white}               &&  0.1874 (0.0435)    \\
\textsc{male}                &&  0.1503 (0.05515    \\
\textsc{white $\times$ male} && -0.2292 (0.0612)   \\
\textsc{college degree}      && 0.2019 (0.0297)    \\
\textsc{employed}            &&  0.2096 (0.0267)    \\
\textsc{age}                 && -0.0091 (0.0008)   \\
\textsc{income 2}            && 0.2803 (.04256)    \\
\textsc{income 3}            && 0.4164 (.0423)     \\
\textsc{income 4}            &&  0.5904 (.04482) \\
\textsc{income 5}            && 0.6241 (.0510)    \\
\\
\textit{Dependent Variable: Sadness}\\
\hline
\textsc{white}               &  &  -0.0706 (0.0460) \\
\textsc{male}                &  &  -0.0310 (0.0582) \\
\textsc{white $\times$ male} &  &  0.1480 (0.0642)  \\
\textsc{college degree}      &  &  0.0332 (0.0302)  \\
\textsc{employed}            &  & 0.0176 (0.02279) \\
\textsc{age}                 &  & -0.0017 (0.0008) \\
\textsc{income 2}            &  & -0.0863 (0.0442) \\
\textsc{income 3}            &  & -0.2384 (0.0444) \\
\textsc{income 4}            &  & -0.3464 (0.0471) \\
\textsc{income 5}            &  &  -0.4734 (0.0535) \\
\textsc{with partner}        &  &  -0.3028 (0.0257) \\
\hline
$\rho$ &&  -0.3814 (0.1266) \\
$N$  && 9110\\
\hline
\hline
\end{tabular} \vspace*{-0.1cm}
\label{tabHEALTH1}
\begin{tablenotes}[flushleft]
\small
\item Notes: Table \ref{tabHEALTH1} reports coefficient estimates from the health and happiness specification.  Standard errors (reported in parentheses)  are Huber/White sandwich estimates.
\end{tablenotes}
\end{threeparttable}
\end{table}

\section{Conclusion}
\label{sec:concl} 

We formulate lattice ordered response models for narrow bracketing, identifying parameters, thresholds, and the joint c.d.f. in a semiparametric framework. In the bivariate probit case, we separately identify $\beta_d$ and $\alpha_{j_d}^{(d)}$ using marginal probabilities, and $\rho$ using joint probabilities with an exclusive covariate. The lattice structure simplifies estimation, suitable for empirical applications. Future work could develop estimation methods.

\bibliography{bibliolattice}

\section*{Appendix}

\textbf{Proof of Theorem \ref{th:semiparamident1}.}
Fix dimension $d$, $d=1, \ldots, D$, for which the condition of this theorem holds. Because of the lattice structure and Assumption \ref{assn:ind} we have for any $x_d \in \mathcal{X}_d$,  
\begin{equation} P(Y^{c_d} \leq y^{(d)}_{j} \, | \, x_d) = F_d \left({\alpha}^{(d)}_{j} - x_d \beta_d \right), \quad j=1, \ldots, M_d.
\label{si_lattice_semiparametric}	
\end{equation}
Take $j\leq M_d-1$ that satisfies conditions of this theorem for this $d$. 

Assumption \ref{assn:rank_ind_semiparametric} guarantees that $P(Y^{c_d} \leq y^{(d)}_{j} | x_d)$ will not be degenerate for $x_d \in S^{(d;j)}$ (in the sense that it will not take values 0 or 1 only). Relation (\ref{si_lattice_semiparametric}) is the basis of the identification strategy. Strict monotonicity of c.d.f. $F_d$ automatically gives us that for two $x_d. \tilde{x}_d \in S^{(d;j)}$, 
$$P(Y^{c_d} \leq y^{(d)}_{j} \, | \, \tilde{x}_d) > P(Y^{c_d} \leq y^{(d)}_{j} \, | \, x_d) \text{ for some } j \quad \iff \quad \tilde{x}_d \beta_d < x_d \beta_d.$$
Thus, the identification is similar to the one in single-index models with a monotone link function (e.g. see \cite*{manski1988} for the statistical independence case or \cite*{manski1985} (Lemma 2) for the proof under large support). Notice that we do not need a large support condition for this result.

Note that the sign of $\beta_{d,1}$ can be identified from varying $x_{d,1}$ within the interval $(\underline{x}_{d,1},\overline{x}_{d,1})$ for $x_{d,-1} \in \widetilde{S}^{(d;j)}_{-1}$. If $P(Y^{c_d} \leq y^{(d)}_{j} \, | \, x_d)$ strictly decreases (increases) when $x_{d,1}$ increases within that interval, then $\beta_{d,1}>0$ ($\beta_{d,1}<0$). If it does not change then $\beta_{d,1}=0$ but this case was ruled by the conditions of the theorem. For concreteness suppose $\beta_{d,1}>0$. Then normalize it as $\beta_{d,1}=1$ to fix the scale. 

Take $b_d \neq \beta_d$ (both are normalized in the same way so $b_{d,1}=1$ and $\beta_{d,1}=1$). 
The conditions of the theorem imply that there exists a positive measure of $x^0_{d,-1} \in \widetilde{S}^{(d;j)}_{ -1}$ such that $x^0_{d,-1} \beta_{-1} \neq x^0_{d,-1} b_{-1}$. Without a loss of generality suppose that for a positive measure of such $x^0_{d,-1}$, we have $x^0_{d,-1} \beta_{-1} > x^0_{d,-1} b_{-1}$.  For any $x_{d,1}^{0}$ that complements  $x^0_{d,-1}$ to a point in $\widetilde{S}^{(d;j)}$ we clearly have  $x_{d,1}^{0}+x^0_{d,-1} \beta_{-1} > x_{d,1}^{0}+x^0_{d,-1} b_{-1}$. We can take $x_{d,1}^{0} \in (\underline{x}_{d,1}, \overline{x}_{d,1})$. 

Due to the continuity of the regressor $x_{d,1}$ on $(\underline{x}_{d,1}, \overline{x}_{d,1})$, one can  find $\tilde{x}^0_{d,1}$ slightly different from $x^0_{d,1}$ such that  $(\tilde{x}^0_{d,1}, x^0_{d,-1})  \in \widetilde{S}^{(d;j)}$ and 
$$x_{d,1}^{0}+x^0_{d,-1} \beta_{d,-1} \stackrel{(*)}{>} \tilde{x}^0_{d,1}+x^0_{d,-1} \beta_{d,-1} \stackrel{(**)}{>} x_{d,1}^{0}+x^0_{d,-1} b_{d,-1}.$$
If $b$ and $\beta$ were \textit{both} consistent with the observables, we would have from  (*) that 
\begin{equation} P\left(Y^{c_d} \leq y^{(d)}_{j} \, | \, (x_{d,1}^{0},x^0_{d,-1}) \right) < P\left( Y^{c_d} \leq y^{(d)}_{j} \, | \, (\tilde{x}_{d,1}^{0},x^0_{d,-1}) \right),
\label{ineq_contr1_lattice}
 \end{equation}
and from inequality (**) that 
\begin{equation}  P\left( Y^{c_d} \leq y^{(d)}_{j} \, | \, (\tilde{x}_{d,1}^{0},x^0_{d,-1}) \right)< P\left(Y^{c_d} \leq y^{(d)}_{j} \, | \, (x_{d,1}^{0},x^0_{d,-1}) \right). 
\label{ineq_contr2_lattice}
\end{equation}
Inequalities (\ref{ineq_contr1_lattice}) and (\ref{ineq_contr2_lattice}) give a contradiction for the probability on the left-hand side of (\ref{ineq_contr1_lattice}). This contradiction is obtained for a positive measure of $(x_{d,1}^{0}, x_{d,-1}^{0})$. This implies that $\beta_d$ is identified relative to $b_d$. $\blacksquare$  

\vskip 0.05in 

\textbf{Proof of Theorem \ref{th:semiparamident2}.}  Fix a dimension $d$, $d=1, \ldots, D$, for which the condition of this theorem holds. Also fix $j=1, \ldots, M_d-2$. Then for $x_d \in S^{(d;j)}$ and $\tilde{x}_d \in S^{(d;j+1)}$ such that $P(Y^{c_d} \leq y^{(d)}_{j} \, | \, x_d ) = P( Y^{c_d} \leq y^{(d)}_{j+1} \, | \, \tilde{x}_{d})$ we have 
$$F_d \left({\alpha}^{(d)}_{j} - x_d \beta_d \right) = F_d \left({\alpha}^{(d)}_{j+1} - \tilde{x}_d \beta_d \right) \in (0,1).$$
Using the convexity of the support of $\varepsilon_d$ in Assumption \ref{assn:ind} and, thus, strict monotonicity of $F_d$ in the interior, we conclude right away that 
${\alpha}^{(d)}_{j+1} - {\alpha}^{(d)}_{j} = \tilde{x}_d \beta_d - x_d \beta_d.$  
Since $\beta_d$ is already identified by Theorem \ref{th:semiparamident1}, we immediately conclude that ${\alpha}^{(d)}_{j+1} - {\alpha}^{(d)}_{j}$ is identified for any $j=1, \ldots, M_d-2$. $\blacksquare$

\vskip 0.05in 

\textbf{Proof of Theorem \ref{corollary:Fd}.} 
(i) Using the result of Theorem \ref{th:semiparamident2}, we conclude that in this case all the thresholds $\alpha^{(d)}_{j}$, $j=1,\ldots, M_d-1$ become known. Then all the underlying components $\alpha^{(d)}_{j}-x_d\beta_d$. From condition (\ref{extracond}), we conclude that known $\alpha^{(d)}_{j}-x_d\beta_d$ cover the whole support of $\varepsilon_d$ (potentially with the choice of different $j$). Hence, known $F_d(\alpha^{(d)}_{j}-x_d\beta_d)$ for known $\alpha^{(d)}_{j}-x_d\beta_d$ identify the marginal c.d.f $F_d$. 
\\(ii) If $F_d(e_{0d})=c_{0d}$, this allows us to find $\alpha^{(d)}_j=x_d\beta_d+e_{0d}$ in the expression where 
$F_d(\alpha^{(d)}_j-\alpha^{(d)}_j)=c_{0d}$. Once one $\alpha^{(d)}_j$ is known, we proceed as in (i).  
$\blacksquare$

\vskip 0.05in 

\textbf{Proof of Theorem \ref{th:semiparamident3}.}  
Consider $$P\left(Y^{(1)} \leq y^{(1)}_{j_1}, \ldots, Y^{(D)} \leq y^{(D)}_{j_D}|x\right) = F(\alpha^{(1)}_{j_1} - x_1\beta_1, \ldots, \alpha^{(D)}_{j_D} - x_D\beta_D).$$ 
For any given value $p_0 \in (0,1)$ of this observed probability, we want to pin down the whole $(D-1)$-dimensional surface of values $\alpha^{(1)}_{j_1} - x_1\beta_1, \ldots, \alpha^{(D)}_{j_D} - x_D\beta_D$ that produce this value. 

Let us identify the pre-image of c.d.f. $F_{12}$ for any $p_0$. Take any value of $x_1$ such that 
$$P\left(Y^{(1)} \leq y^{(1)}_{j_1}|x_1\right)=\tilde{p}_0>p_0.$$ 
Now we operate with processes that do have exclusive covariates. Start by varying exclusive covariate $x_{2,1}$. By the conditions of this theorem, by changing $x_{2,1}$ alone we can force the choice probability 
$$P\left(Y^{(1)} \leq y^{(1)}_{j_1}), Y^{(2)} \leq y^{(2)}_{j_2})|x_1, x_2\right)$$ 
to vary from 0 to $\tilde{p}_0$. We will consider only that variation that makes this probability $p_0$. This will identify the pre-image of the c.d.f  $F_{12}$ corresponding to $p_0$.  By taking $p_0$ arbitrary in $(0,1)$ we effectively identify the joint c.d.f $F_{12}$.

Let us identify the pre-image of $F_{123}$ for any $p_0$. Take any value of $x_1$ and $x_2$ such that 
$$P\left(Y^{(1)} \leq y^{(1)}_{j_1}, Y^{(2)} \leq y^{(2)}_{j_2}|x_1, x_2\right)=\tilde{p}_0>p_0.$$  Now vary the exclusive covariate $x_{3,1}$.  By the conditions of this theorem, by changing $x_{3,1}$ and  we can force the choice probability 
$$P\left(Y^{(1)} \leq y^{(1)}_{j_1}), Y^{(2)} \leq y^{(2)}_{j_{2}}), Y^{(3)} \leq y^{(3)}_{j_{3}})|x_1, x_{2},x_{3}\right)$$ 
to vary from 0 to $\tilde{p}_0$. We pick only those directions that make this probability $p_0$. This will identify the pre-image of the c.d.f  $F_{123}$ corresponding to $p_0$.  By taking $p_0$ arbitrary in $(0,1)$ we effectively identify the joint c.d.f $F_{123}$. 

Proceeding sequentially in this manner, we identify the overall joint c.d.f. $F$.  $\blacksquare$

\vskip 0.05in

\textbf{Proof of Theorem \ref{th:probit_index}.}
Using the lattice assumption, 
$P(Y^{(d)}\le y^{(d)}_j \mid x_d)=\Phi(\alpha^{(d)}_j - x_d\beta_d).$ Apply $\Phi^{-1}$ to observed conditional probabilities to obtain linear equations of the form $\Phi^{-1}\left(P(Y^{(d)}\le y^{(d)}_j \mid x_d) \right)=\alpha^{(d)}_j - x_d\beta_d$. With $k_d+1$ distinct $x_d$ points and full rank the linear system identifies $\alpha^{(d)}_j$, $j=1,\ldots, M_d-1$, and $\beta_d$. $\blacksquare$

\vskip 0.05in 

\textbf{Proof of Theorem \ref{th:rho_id}.}
(a) Since $P(Y^{c_1}\leq y^{(d_1)}_{j_1} |x^*_{d_1})=\Phi\left( \alpha^{(d_1)}_{j_1}-x^*_{d_1}\beta_{d_1}\right)$, the condition of this part means  that $\alpha^{(d_1)}_{j_1}-x^*_{d_1}\beta_{d_1}=0$. Find the whole vector $x^*$ that has $x^*_{d_1}$ as a vector of covariates in the $d_1$-th process, and extract $x^*_{d_2}$ from $x^*$. If we can find $j_2$ such that  $\alpha^{(d_2)}_{j_2}-x^*_{d_2}\beta_{d_2} \leq 0$, then we consider the observed  probability   
$$
P\left(Y^{(d_1)} \leq y^{(d_1)}_{j_1}, \;  Y^{(d_2)} \leq y^{(d_2)}_{j_2}\, | \, x^*_{d_1}, x^*_{d_2} \right)  = \int_{-\infty}^{\alpha^{(d_2)}_{j_2}-x^*_{d_2}\beta_{d_2}} \dfrac{1}{\sqrt{2\pi}}e^{-\frac{\eta^2}{2}} \Phi\left(-\frac{\rho_{d_1,d_2}}{\sqrt{1-\rho_{d_1,d_2}^2}} \eta \right) d \eta.  
$$
Because $\alpha^{(d_2)}_{j_2}-x^*_{d_2}\beta_{d_2} \leq 0$,  the right-hand side is strictly increasing in $\frac{\rho_{d_1,d_2}}{\sqrt{1-\rho_{d_1,d_2}^2}}$  and everything else on the right-hand side is known. Therefore, $\frac{\rho_{d_1,d_2}}{\sqrt{1-\rho_{d_1,d_2}^2}}$ is identified. Since $\frac{\rho_{d_1,d_2}}{\sqrt{1-\rho_{d_1,d_2}^2}}$ in its turn is a strictly increasing function of $\rho_{d_1,d_2} \in (-1,1)$, this guarantees that identification of $\rho_{d_1,d_2}$. 

If $\alpha^{(d_2)}_{j_2}-x^*_{d_2}\beta_{d_2}< 0$ for any $j_2$, then instead we would consider the probability $P\left(Y^{(d_1)} \leq y^{(d_1)}_{j_1}, \;  Y^{(d_2)} > y^{(d_2)}_{j_2} \, | \, x^*_{d_1}, x^*_{d_2}  \right)$ and conduct an analogous identification strategy.

(b) The first inequality implies that $\alpha^{(d_2)}_{j_2}-x_{d_2}\beta_{d_2}$ and $\alpha^{(d_2)}_{j_2}-\widetilde{x}_{d_2}\beta_{d_2}$ have the same sign and the second inequality implies that this sign is opposite to the sign of  $\alpha^{(d_2)}_{j_2}-{x}^{\diamond}_{d_2}\beta_{d_2}$. For concreteness suppose that the first two expressions are positive and the third one is negative. 

The third inequality implies that $\alpha^{(d_1)}_{j_1}-x_{d_1}\beta_{d_1}$ and $\alpha^{(d_1)}_{j_1}-\widetilde{x}_{d_1}\beta_{d_1}$ have different signs. Without a loss of generality,  $\alpha^{(d_1)}_{j_1}-x_{d_1}\beta_{d_1} >0$ and $\alpha^{(d_1)}_{j_1}-\widetilde{x}_{d_1}\beta_{d_1}<0$. 

Consider 
$$
	P\left(Y^{(d_1)} \leq y^{(d_1)}_{j_1},  Y^{(d_2)} > y^{(d_2)}_{j_2} \, | \, x \right)  = \int_{\alpha^{(d_2)}_{j_2}	-x_{d_2}\beta_{d_2}}^{+\infty} \dfrac{1}{\sqrt{2\pi}}e^{-\frac{\eta^2}{2}} \Phi\left(\frac{\alpha^{(d_1)}_{j_1}	-x_{d_1}\beta_{d_1}-\rho_{d_1,d_2} \eta}{\sqrt{1-\rho_{d_1,d_2}^2}} \right) d \eta, 
	$$
    where the only unknown on the right-hand side is $\rho_{d_1,d_2}$ and $-\frac{\rho_{d_1,d_2} \eta}{\sqrt{1-\rho_{d_1,d_2}^2}}$ is strictly decreasing in $\rho_{d_1,d_2}$. Since $\alpha^{(d_1)}_{j_1}-x_{d_1}\beta_{d_1} >0 0$, then $\frac{\alpha^{(d_1)}_{j_1^0}-x^{(1)}_{d_1}\beta_{d_1}}{\sqrt{1-\rho_{d_1,d_2}^2}}$ as a function of $\rho_{d_1,d_2}$ is  decreasing on the interval $(-1,0]$. Hence, the whole right-hand of this probability expression is strictly decreasing in $\rho_{d_1,d_2}$ on the interval $\left(-1,0\right]$. Thus,  among non-positive $\rho_{d_1,d_2}$, there can be at most one value that can generate observable left-hand side.

    Consider 
{\small$$
	P\left(Y^{(d_1)} \leq y^{(d_1)}_{j_1},  Y^{(d_2)} > y^{(d_2)}_{j_2} \, | \, \widetilde{x} \right)  = \int_{\alpha^{(d_2)}_{j_2}	-\widetilde{x}_{d_2}\beta_{d_2}}^{+\infty} \dfrac{1}{\sqrt{2\pi}}e^{-\frac{\eta^2}{2}} \Phi\left(\frac{\alpha^{(d_1)}_{j_1}	-\widetilde{x}_{d_1}\beta_{d_1}-\rho_{d_1,d_2} \eta}{\sqrt{1-\rho_{d_1,d_2}^2}} \right) d \eta, 
	$$}
     Since $\alpha^{(d_1)}_{j_1}-\widetilde{x} _{d_1}\beta_{d_1} <0 $, then the right-hand side of the last equation is strictly decreasing in $\rho_{d_1,d_2}$ on the interval $[0,1)$. Hence, among non-negative $\rho_{d_1,d_2}$, there can be at most one value that can generate observables. 
     
     Thus, at this stage of the proof there can be at most two values (one non-negative and one non-positive) in the identified set. Let us denote these two candidate values as $\rho_{d_1,d_2}^* \leq 0$ and $\tilde{\rho}_{d_1,d_2}>0$.

     Now consider 
	{\small$$
	P\left(Y^{(d_1)} \leq y^{(d_1)}_{j_1},  Y^{(d_2)} \leq y^{(d_2)}_{j_2} \, | \, x^{\diamond} \right)  = \int_{-\infty}^{\alpha^{(d_2)}_{j_2}	-x^{\diamond}_{d_2}\beta_{d_2}} \dfrac{1}{\sqrt{2\pi}}e^{-\frac{\eta^2}{2}} \Phi\left(\frac{\alpha^{(d_1)}_{j_1}	-x^{\diamond}_{d_1}\beta_{d_1}-\rho_{d_1,d_2} \eta}{\sqrt{1-\rho_{d_1,d_2}^2}} \right) d \eta.  \footnote{The reason we consider $Y^{(d_2)} \leq y^{(d_2)}_{j_2}$ is because $\alpha^{(d_2)}_{j_2}	-{x}^{\diamond}_{d_2}\beta_{d_2}<0$.}
	$$}
	Since $\alpha^{(d_2)}_{j_2}	-x^{\diamond}_{d_2}\beta_{d_2} <0 $, the equation 
	$$P\left(Y^{(d_1)} \leq y^{(d_1)}_{j_1},  Y^{(d_2)} \leq y^{(d_2)}_{j_2} \, | \, x^{\diamond} \right)  = \int_{-\infty}^{\alpha^{(d_2)}_{j_2}	-x^{\diamond}_{d_2}\beta_{d_2}} \dfrac{1}{\sqrt{2\pi}}e^{-\frac{\eta^2}{2}} \Phi\left(b-a \eta \right)  d \eta 
	$$
considered for all observationally equivalent $(a,b)$, delivers a strictly decreasing in $a$ function $b(a)$ that generates the same $P\left(Y^{(d_1)} \leq y^{(d_1)}_{j_1},  Y^{(d_2)} \leq y^{(d_2)}_{j_2} \, | \, x^{\diamond}\right)$. It is easy to see that for both $a^*=\frac{\rho_{d_1,d_2}^*}{\sqrt{1-{\rho^*_{d_1,d_2}}^2}}\leq 0$, $b^*=\frac{\alpha^{(d_1)}_{j_1^0}	-x^{\diamond}_{d_1}\beta_{d_1}}{\sqrt{1-{\rho^*_{d_1,d_2}}^2}}<0 $ and  $\tilde{a}=\frac{\tilde{\rho}_{d_1,d_2}}{\sqrt{1-\tilde{\rho}_{d_1,d_2}^2}}> 0$, $\tilde{b}=\frac{\alpha^{(d_1)}_{j_1^0}	-x^{\diamond}_{d_1}\beta_{d_1}}{\sqrt{1-\tilde{\rho}_{d_1,d_2}^2}}<0 $ to be compatible with the fact that they belong long to the curve $(a,b(a))$ with the strictly decreasing $b(\cdot)$, it has to be satisfied that  $|\tilde{\rho}_{d_1,d_2}| >|{\rho}^*_{d_1,d_2}|$.

Going back to $\widetilde{x}$ note that since $\alpha^{(d_1)}_{j_1}	-\widetilde{x}^{(2)}_{d_1}\beta_{d_1} <0 $, the equation 
	{\small$$P\left(Y^{(d_1)} \leq y^{(d_1)}_{j_1},  Y^{(d_2)} \leq y^{(d_2)}_{j_2} \, | \, \widetilde{x} \right)  = \int_{-\infty}^{\alpha^{(d_1)}_{j_1}	-\widetilde{x}_{d_1}\beta_{d_1}} \dfrac{1}{\sqrt{2\pi}}e^{-\frac{\eta^2}{2}} \Phi\left(b-a \eta \right)  d \eta  
	$$}considered for all observationally equivalent $(a,b)$, delivers a strictly decreasing in $a$ function $b(a)$ that generates the same $P\left(Y^{(d_1)} \leq y^{(d_1)}_{j_1},  Y^{(d_2)} \leq y^{(d_2)}_{j_2} \, | \, \widetilde{x}\right)$. It is easy to see that for both $a^*=\frac{\rho_{d_1,d_2}^*}{\sqrt{1-{\rho^*_{d_1,d_2}}^2}}\leq 0$, $b^*=\frac{\alpha^{(d_1)}_{j_1}	-\widetilde{x}_{d_1}\beta_{d_1}}{\sqrt{1-{\rho^*_{d_1,d_2}}^2}}>0 $ and  $\tilde{a}=\frac{\tilde{\rho}_{d_1,d_2}}{\sqrt{1-\tilde{\rho}_{d_1,d_2}^2}}> 0$, $\tilde{b}=\frac{\alpha^{(d_1)}_{j_1}	-\widetilde{x}_{d_1}\beta_{d_1}}{\sqrt{1-\tilde{\rho}_{d_1,d_2}^2}}>0 $ to be compatible with the fact that they belong long to the curve $(a,b(a))$ with the strictly decreasing $b(\cdot)$, it has to be satisfied that $|\tilde{\rho}_{d_1,d_2}| <|{\rho}^*_{d_1,d_2}|$. This is a contradiction with the previous conclusion. Therefore, only one of $\rho_{d_1,d_2}^*$ and $\tilde{\rho}_{d_1,d_2}$ can generate observables.
    
(c)  Denote $x^{(1)}_{d_1}=(x_{d_1,1:L_1}^{(1)},x_{d_1,L_{d_1}+1:k_{d_1}})$ and $x^{(2)}_{d_1}=(x_{d_1,1:L_1}^{(2)},x_{d_1,L_{d_1}+1:k_{d_1}})$.

We first consider the case when 
$\alpha^{(d_1)}_{j_1} - x_{d_1}^{(1)}\beta_{d_1}$ and $\alpha^{(d_1)}_{j_1} - x_{d_1}^{(2)}\beta_{d_1} $ take different signs -- e.g. suppose that $\alpha^{(d_1)}_{j_1} - x_{d_1}^{(1)}\beta_{d_1} \geq 0$ and  $\alpha^{(d_1)}_{j_1} - x_{d_1}^{(2)}\beta_{d_1} \leq 0$. 

For indices $j_1$ and $j_2$ in condition in (c),  consider the probability  
	\begin{equation} 
		\label{eq:case22_C2} P\left(Y^{c_1} \leq y^{(d_1)}_{j_1},  Y^{c_2} \leq y^{(d_2)}_{j_2} \, | \, x^{(2)}_{d_1}, x_{d_2} \right)  = \int_{-\infty}^{\alpha^{(d_1)}_{j_1} - x_{d_1}^{(2)}\beta_{d_1} }\dfrac{1}{\sqrt{2\pi}}e^{-\frac{\eta^2}{2}} \Phi\left(b -a \eta\right) d \eta, 
	\end{equation}	
	where $a=\frac{\rho_{d_1,d_2}}{\sqrt{1-\rho_{d_1,d_2}^2}}$, $b=\frac{\alpha^{(d_2)}_{j_2} -x_{d_2} \beta_{d_2}}{\sqrt{1-\rho_{d_1,d_2}^2}}$. Because $\alpha^{(d_1)}_{j_{d_1}} - x_{d_1}^{(2)}\beta_{d_1}\leq 0$, the right-hand side of (\ref{eq:case22_C2}) is strictly increasing in $a$. It is obviously also strictly increasing in $b$. This means that for any feasible $a \in \mathbb{R}$ we can find $b_2(a)$ such that 
$$ P\left(Y^{c_1} \leq y^{(d_1)}_{j_1},  Y^{c_2} \leq y^{(d_2)}_{j_2} \, | \, x^{(2)}_{d_1}, x_{d_2} \right) \\ = \int_{-\infty}^{\alpha^{(d_1)}_{j_1} - x_{d_1}^{(2)}\beta_{d_1} }\dfrac{1}{\sqrt{2\pi}}e^{-\frac{\eta^2}{2}} \Phi\left(b_2(a) -a \eta\right) d \eta, 
$$	
and $b_2(\cdot) $ is a strictly decreasing function. Now consider the probability  
$$
P\left(Y^{c_1} > y^{(d_1)}_{j_1},  Y^{c_2} \leq y^{(d_2)}_{j_2} \, | \, x^{(1)}_{d_1}, x_{d_2} \right)  = \int_{\alpha^{(d_1)}_{j_1} - x_{d_1}^{(1)}\beta_{d_1} }^{+\infty} \dfrac{1}{\sqrt{2\pi}}e^{-\frac{\eta^2}{2}} \Phi\left(b -a \eta\right) d \eta, 
$$ 
	where $a$ and $b$ are the same as in (\ref{eq:case22_C2}). Because $\alpha^{(d_1)}_{j_1} - x_{d_1}^{(1)}\beta_{d_1} \geq 0$, the right-hand side of the last expression is strictly decreasing in $a$. It is obviously also strictly increasing in $b$. This means that for any  feasible $a \in \mathbb{R}$ we can find $b_1(a)$ such that 
	$$
		 P\left(Y^{c_1} > y^{(d_1)}_{j_1},  Y^{c_2} \leq y^{(d_2)}_{j_2} \, | \, x^{(1)}_{d_1}, x_{d_2} \right)  = \int_{\alpha^{(d_1)}_{j_1} - x_{d_1}^{(1)}\beta_{d_1}}^{+\infty} \dfrac{1}{\sqrt{2\pi}}e^{-\frac{\eta^2}{2}} \Phi\left(b_1(a) -a \eta\right) d \eta.
$$
Note that since we only vary the first $L_{d_1}$ covariates in $x_{d_1}$, which are excluded from  $x_{d_2}$, then $\alpha^{d_2}_{j_2} -x_{d_2} \beta_{d_2}$ does not vary. This implies  that $\rho_{d_1,d_2}$ is identified because the strictly increasing $b_1(\cdot)$ and the strictly decreasing $b_2(\cdot)$ can intersect only once and the argument at that intersection is at $\frac{\rho_{d_1,d_2}}{\sqrt{1-\rho_{d_1,d_2}^2}}$, which can be inverted to give $\rho_{d_1,d_2}$.

\begin{figure}
\centering
    \caption{Functions $b_2(\cdot)$ (solid line) and $b_1(\cdot)$ (dotted line)}	
    \includegraphics[width=0.4\textwidth]{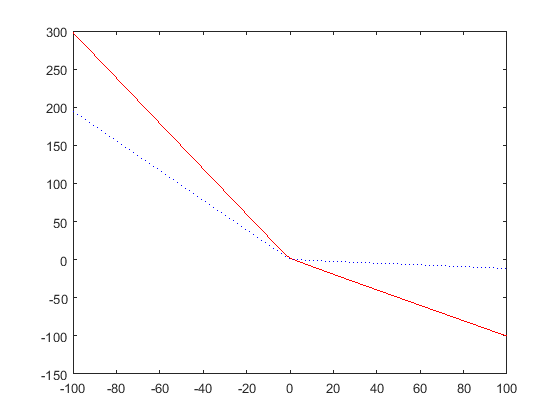}
	    \label{fig:normal_twocurves1}
\end{figure}

We now consider the case when both 
$\alpha^{(d_1)}_{j_1} - x_{d_1}^{(1)}\beta_{d_1}$ and $\alpha^{d_1}_{j_1} - x_{d_1}^{(2)}\beta_{d_1}$ have the same sign. Suppose that they are both non-positive.\footnote{If they are both non-negative, then instead of considering the conditional probabilities of $\{Y^{c_1} \leq y^{(d_1)}_{j_1},  Y^{c_2} \leq y^{(d_2)}_{j_2} \}$ we would consider the conditional probabilities of $\{Y^{c_1} > y^{(d_1)}_{j_1},  Y^{c_2} \leq y^{(d_2)}_{j_2} \}$.} Without a loss of generality,  
$$P\left(Y^{c_1} \leq y^{(d_1)}_{j_1},  Y^{c_2} \leq y^{(d_2)}_{j_2} \, | \, x_{d_1}^{(1)},  x_{d_2} \right) 
	>  \\ 
	P\left(Y^{(d_1)} \leq y^{(d_1)}_{j_1},  Y^{(d_2)} \leq y^{(d_2)}_{j_2} \, | \, x_{d_1}^{(2)}, x_{d_2} \right).$$
Then both level functions $b_2(\cdot)$ and $b_1(\cdot)$ defined by equations 
$$
P\left(Y^{c_1} \leq y^{(d_1)}_{j_1},  Y^{c_2} \leq y^{(d_2)}_{j_2} \, | \, x^{(2)}_{d_1}, x_{d_2} \right)  = \int_{-\infty}^{\alpha^{(d_1)}_{j_1} - x_{d_1}^{(2)}\beta_{d_1}}\dfrac{1}{\sqrt{2\pi}}e^{-\frac{\eta^2}{2}} \Phi\left(b_1(a) -a \eta\right) d \eta
$$
	and 
$$	
P\left(Y^{c_1} \leq y^{(d_1)}_{j_1},  Y^{c_2} \leq y^{(d_2)}_{j_2} \, | \, x^{(2)}_{d_1}, x_{d_2} \right) \\ = \int_{-\infty}^{\alpha^{(d_1)}_{j_1} - x_{d_1}^{(1)}\beta_{d_1}}\dfrac{1}{\sqrt{2\pi}}e^{-\frac{\eta^2}{2}} \Phi\left(b_2(a) -a \eta\right) d \eta
$$	
are strictly decreasing. However, the function $b_1(a)$ has a derivative that is strictly greater than the derivative of $b_2(a)$ for all $a$ in the intersection of feasible sets. Moreover, for all low enough common feasible $a$ the values of $b_1(a)$ are lower than the values of $b_2(a)$ and for all high enough $a$ the values of $b_1(a)$ are higher than the values of $b_2(a)$. This situation is illustrated in Figure \ref{fig:normal_twocurves1}. Together with the strict inequality on the derivatives of these functions, these properties imply that these two functions may intersect only once. Their intersection is at $\frac{\rho_{d_1,d_2}}{\sqrt{1-\rho_{d_1,d_2}^2}}$, which can be inverted to give $\rho_{d_1,d_2}$. 
    $\blacksquare$

\end{document}